\documentclass[12pt]{article}
\usepackage{amssymb}

\usepackage{graphicx}
\usepackage{amsmath}

\newtheorem{theorem}{Theorem}
\newtheorem{acknowledgement}[theorem]{Acknowledgement}

\input{tcilatex}

\begin{document}

\title{{\Large Canonical form of Euler-Lagrange equations and gauge
symmetries}}
\author{B. Geyer\thanks{%
Naturwissenschaftlich-Theoretisches Zentrum und Institut f\"{u}r
Theoretische Physik, Universit\"{a}t Leipzig, Germany; e-mail:
geyer@itp.uni-leipzig.de}, \ D.M. Gitman\thanks{%
Institute of Physics, University of Sao Paulo, Brazil; e-mail:
gitman@fma.if.usp.br}, \ and I.V. Tyutin\thanks{%
Lebedev Physics Institute, Moscow, Russia; e-mail: tyutin@lpi.ru}}
\date{\today            }
\maketitle

\begin{abstract}
The structure of the Euler-Lagrange equations for a general Lagrangian
theory (e.g. singular, with higher derivatives) is studied. For these
equations we present a reduction procedure to the so-called canonical form.
In the canonical form the equations are solved with respect to highest-order
derivatives of nongauge coordinates, whereas gauge coordinates and their
derivatives enter in the right hand sides of the equations as arbitrary
functions of time. The reduction procedure reveals constraints in the
Lagrangian formulation of singular systems and, in that respect, is similar
to the Dirac procedure in the Hamiltonian formulation. Moreover, the
reduction procedure allows one to reveal the gauge identities between the
Euler-Lagrange equations. Thus, a constructive way of finding all the gauge
generators within the Lagrangian formulation is presented. At the same time,
it is proven that for local theories all the gauge generators are local in
time operators.
\end{abstract}

\section{Introduction}

At present increasingly complicated gauge models are used in field and
string theory. Generally a comprehensive analysis of their structure is not
a simple task. In the Lagrangian formulation the problem includes, in
particular, finding generators of gauge symmetries and their algebra,
revealing the hidden structure of the equations of motion and so on. One
ought to say that in the Hamiltonian formulation there exists a relatively
well-developed scheme of constraint finding (Dirac procedure \cite{Dirac64})
and reorganization \cite{Dirac64,GitTy90,HenTe92,GitTy01}. The constraint
structure can be, in principle, related to the symmetry properties of the
initial gauge theory in the Lagrangian formulation \cite{BorTy98}. However,
in the general case, this relation cannot be considered as a constructive
method to study the Lagrangian symmetries (it is indirect and complicated).
Moreover, the modern tendency is to avoid the non-covariant hamiltonization
step and to use the Lagrangian quantization \cite{BV} for constructing
quantum theory. Such an approach incorporates all the Lagrangian structures
(in particular, the total gauge algebra).{\LARGE \ }That is why it seems
important to develop a reduction procedure within the Lagrangian formulation
-- in a sense similar to the Dirac procedure in the Hamiltonian formulation
-- that may allow one in a constructive manner to reveal the hidden
structure of the Euler-Lagrange equations (ELE) of motion and to find all
the gauge identities and therefore the generators of all the gauge
transformations. An idea of such a procedure was first mentioned in
publications of the authors (D.G and I.T) \cite{GitTy83,GitTy87} (see also
Appendix C in \cite{GitTy90}), but was not appropriately\ elaborated and
some important points where not revealed.

In the present paper we return to this idea studying the structure of the
ELE for a general Lagrangian theory (singular, with higher derivatives, and
with external fields). In Sect. II we introduce some notation and
definitions. In Sect. III, we reduce the ELE of nonsingular theories to the
so called canonical form (in the canonical form the equations are solved
with respect to highest-order derivatives of nongauge coordinates, whereas,
gauge coordinates and their derivatives enter in the right hand sides of the
equations as arbitrary functions of time, see below). In Sect. IV we
formulate the\ reduction procedure for the singular case. In a sense, the
reduction procedure reveals constraints in the Lagrangian formulation of
singular systems and, in that respect, is similar to the Dirac procedure in
the Hamiltonian formulation. In Sect. V we demonstrate how the reduction
procedure reveals the gauge identities between the ELE. Thus, a constructive
way of finding all the gauge generators within the Lagrangian formulation is
presented. At the same time it is proven that for local theories all the
gauge generators are local in time operators. In the Appendix we collect
some Lemmas being useful for our consideration.

\section{General ELE}

\subsection{Notation, definitions, and conventions}

We consider a system with finite degrees of freedom (classical mechanics).
These degrees of freedom are described by the generalized coordinates $%
q^{a}, $ $a=1,...,n,$ which depend on the time $t.$ The following notation
is used:%
\begin{equation}
\dot{q}^{a}=\frac{dq^{a}}{dt}\,,\;\ddot{q}^{a}=\frac{d^{2}q^{a}}{dt^{2}}%
\,,\;\cdots \,,\;\;\mathrm{or\;\;}q^{a\left[ l\right] }=\frac{d^{l}q^{a}}{%
dt^{l}}\,,\;\;l=0,1,...,\;\;\left( q^{a\left[ 0\right] }=q^{a}\right) \,.
\label{2.}
\end{equation}%
The coordinates $q^{a}=q^{a\left[ 0\right] }$ are called sometimes
velocities of zeroth order; the velocities $\dot{q}^{a}=q^{a\left[ 1\right]
} $ are called velocities of the first order; the accelerations $\ddot{q}%
^{a}=q^{a\left[ 2\right] }$ are called velocities of second order,$\,$and so
on. The space of all the velocities is often called the jet space, see \cite%
{Kuper79}.

As local functions (LF) we call those functions that are defined on the jet
space and depend on the velocities $q^{a\left[ l\right] }\;$up to some
finite orders $N_{a}\geq 0\;(l\leq N_{a})$. Further, we call $N_{a}$ the\
order of the coordinate $q^{a}$ in the LF.\ For the LF we use the following
notation\footnote{%
The functions $F$ may depend on time explicitly, however, we do not include $%
t$ in the arguments of the functions.}:%
\begin{eqnarray}
&&F\left( q^{a},\dot{q}^{a},\ddot{q}^{a},...\right) =F\left( q^{a\left[ 0%
\right] },q^{a\left[ 1\right] },q^{a\left[ 2\right] },...\right) =F\left( q^{%
\left[ l\right] }\right) \,,\;q^{\left[ l\right] }=(q^{a\left[ l\right]
},\;0\leq l\leq N_{a}),  \notag \\
&&\mathrm{or\;sometimes}:\;F\left( q^{\left[ l\right] }\right) =F\left(
\cdots q^{a\left[ N_{a}\right] }\right) \,.  \label{2.a}
\end{eqnarray}%
In the latter form, we indicate only the highest-order derivatives in the
arguments of the LF.

The following notation is often used: $\left[ a\right] $ is the number of
the indices $a,$ namely, if $\ a=1,...n,$ then \ $\left[ a\right] =n.$
Similarly, suppose $F_{a}\left( \eta \right) ,\;a=1,...,n$ are some
functions, then $\left[ F\right] $ is the number of these functions, $\left[
F\right] =n$, etc. . However differently, writing $q^{a\left[ l\right] }$ we
denote by $\left[ l\right] $ the order of the time derivatives, see (\ref{2.}%
).

On the jet space, we define local operators (LO) to be matrix operators $%
\hat{U}$ of the form%
\begin{equation}
\hat{U}_{Aa}=\sum_{k=0}^{K}u_{Aa}^{k}\left( \frac{d}{dt}\right) ^{k}\,,
\label{2.10}
\end{equation}%
where $K$ is a finite number and $u_{Aa}^{k}$ are some LF. The LO act on
columns of LF $f_{a}$ producing columns of LF $F_{A}=\hat{U}_{Aa}f_{a}\,$as
well. We define the transposed operator\emph{\ }to $\hat{U}$ as%
\begin{equation}
\left( \hat{U}^{T}\right) _{aA}=\sum_{k=0}^{K}\left( -\frac{d}{dt}\right)
^{k}u_{Aa}^{k}=\sum_{k=0}^{K}\left( -1\right) ^{k}\sum_{l=0}^{k}\binom{k}{l}%
u_{Aa}^{k\left[ l\right] }\left( \frac{d}{dt}\right) ^{k-l}\,.  \label{2.11}
\end{equation}%
Then the following relation holds true%
\begin{equation}
F_{A}\hat{U}_{Aa}f_{a}=\left[ \left( \hat{U}^{T}\right) _{aA}F_{A}\right]
f_{a}+\frac{d}{dt}Q\,,  \label{2.12}
\end{equation}%
where $F_{A}$ , $f_{a}$\thinspace , and $Q$ are LF. The LO $\hat{U}_{ab}$ is
symmetric ($+$) or skewsymmetric ($-$) whenever the relation $\left( \hat{U}%
^{T}\right) _{ab}=\pm \hat{U}_{ab}\,$ holds true.

Suppose a set of LF $F_{A}\left( \cdots q^{a\left[ N_{a}^{A}\right] }\right) 
$, or a set of equations $F_{A}\left( \cdots q^{a\left[ N_{a}^{A}\right]
}\right) =0\,$, be given. In the general case the orders $N_{a}^{A}\ $of the
coordinates $q^{a}$ in the functions $F_{A}$ (in the equations $F_{A}=0$)
are different, i.e. these orders depend both on $a$ and $A$. The number $%
\mathcal{N}_{a}=\max_{A}N_{a}^{A}$ is called the order of the coordinate $%
q^{a}$ in the set of the functions $F_{A}$ (in the set of the equations $%
F_{A}=0$).

Whenever, a subset $F_{A^{\prime }}=0,\;A^{\prime }\subset A$ has orders $%
\mathcal{N}_{a}^{\prime }$ of the coordinates less than the corresponding
orders of the complete set, namely, $\forall a:$ $\mathcal{N}_{a}^{\prime }<%
\mathcal{N}_{a}$ , we call this subset the constraint equations.

Generally two sets of equations, $F_{A}\left( q^{\left[ l\right] }\right) =0$
and $f_{\alpha }\left( q^{\left[ l\right] }\right) =0$ are equivalent
whenever they have the same set of solutions. In what follows we denote this
fact as: $F=0\Longleftrightarrow f=0.$

Suppose that two sets of LF $F_{A}\left( q^{\left[ l\right] }\right) $ and $%
\chi _{A}\left( q^{\left[ l\right] }\right) $ , $\left[ F\right] =\left[
\chi \right] \,$, are related by some LO,%
\begin{equation}
F=\hat{U}\chi \,,\;\chi =\hat{V}F\,,\;\hat{U}\hat{V}=1\,.  \label{2.9}
\end{equation}%
Then we call such functions equivalent and denote this fact as: $F\sim \chi $
. Obviously,%
\begin{equation}
F\sim \chi \Longrightarrow F=0\Longleftrightarrow \chi =0\,.  \label{2.9c}
\end{equation}%
If (\ref{2.9c}) holds true, we will call the equations $F_{A}=0$ and $%
f_{\alpha }=0$ strong equivalent.

In what follows we often meet the case where%
\begin{equation}
\chi _{A}=\left( 
\begin{array}{c}
f_{\alpha } \\ 
0_{G}%
\end{array}%
\right) ,\;A=\left( \alpha ,G\right) \,;\;\forall G:\,0_{G}\equiv 0\,.
\label{2.9a}
\end{equation}%
Here the equivalence $F\sim \chi $ implies the equivalence of the equations $%
F=0$ and $f=0$ and the existence of the identities $\hat{V}_{GA}F_{A}\equiv
0.$ Namely,%
\begin{equation}
F\sim \chi \,\Longrightarrow \left\{ 
\begin{array}{c}
F=0\Longleftrightarrow f=0 \\ 
\hat{V}_{GA}F_{A}\equiv 0%
\end{array}%
\right. \,.  \label{2.9b}
\end{equation}

\subsection{ELE}

Below we restrict our consideration to the Lagrange functions $L$ that are
LF on the jet space, and depend on some external coordinates (fields) $%
u^{\mu }$ (we call the coordinates $u^{\mu }$ external ones in contrast to
the coordinates $q^{a}$ , which we call inner coordinates) which are some
given functions of time. Thus, 
\begin{equation}
L=L\left( \cdots q^{a\left[ N_{a}\right] };u^{\mu }\right)
\,,\;\;\,a=1,...,n,\;\;N_{a}\geq 0.  \label{2.1}
\end{equation}%
The orders $N_{a}$ of the inner coordinates $q^{a}$ in the Lagrange function
will be called further the proper orders of the coordinates. Coordinates $%
q^{a}$ with the proper orders $N_{a}=0$, we call the degenerate coordinates %
\cite{GitTy02}.

Equations of motion of a Lagrangian theory (the ELE) follow from the action
principle $\delta S=0,$ $\ S=\int Ldt\,$, and have the form (merely the
inner coordinates have to be varied):%
\begin{equation}
\frac{\delta S}{\delta q^{a}}=\sum_{l=0}^{N_{a}}\left( -\frac{d}{dt}\right)
^{l}\frac{\partial L}{\partial q^{a\left[ l\right] }}=0\,,\;\;a=1,...,n\,.
\label{2.2}
\end{equation}

Following \cite{GitTy02}, we classify the Lagrangian theories as nonsingular
($M\neq 0$) and singular ($M=0$) ones by the help of the\ generalized
Hessian $M=\det \left| \left| M_{a\,b}\right| \right| $, where%
\begin{equation}
M_{a\,b}=\frac{\partial ^{2}L}{\partial q^{a\left[ N_{a}\right] }\partial
q^{b\left[ N_{b}\right] }}\,\,  \label{2.3}
\end{equation}%
is the generalized Hessian matrix.

In what follows the ELE of a nonsingular (singular) theory will be called
the\ nonsingular (singular) ELE.

Sometimes, it is convenient to enumerate the inner coordinates and organize
them into groups such that $q^{a}=\left( q^{a_{0}},...,q^{a_{I}}\right) \,,$%
\ where $a_{i}$ are groups of indices that enumerate coordinates having the
same proper orders, $N_{a_{k}}=n_{k}\,$.\ Besides, we organize these groups
such that $n_{I}>n_{I-1}\cdots >n_{0}=0$ $\ (\max \,N_{a}=N_{a_{I}}=n_{I}\,$%
, and $q^{a_{0}}$ are the degenerate coordinates, $N_{a_{0}}=n_{0}=0\,).$
Thus,%
\begin{equation}
a=\left( a_{k}\,,\;k=0,1,...,I\right) \,,\ \;\left[ a\right] =\sum_{i}\left[
a_{i}\right] \,,\;\ \left[ a_{i}\right] \geq 0\,,\;n_{I}>n_{I-1}\cdots
>n_{0}=0\,.  \label{2.3a}
\end{equation}%
Taking into account the notation (\ref{2.3a}), we may write the Lagrange
function and the ELE as: 
\begin{eqnarray}
&&L=L\left( \cdots q^{a_{k}\left[ n_{k}\right] };u^{\mu }\right)
\,,\;\;k=0,1,...,I\,;  \label{2.4} \\
&&F_{a_{k}}\left( \cdots q^{b\left[ N_{b}+n_{k}\right] };\cdots u^{\mu \left[
n_{k}\right] }\right) =0\,,  \label{2.5} \\
&&F_{a_{k}}=\left\{ 
\begin{array}{c}
M_{a_{k}\,b}q^{b\left[ N_{b}+n_{k}\right] }+K_{a_{k}}\left( \cdots q^{b\left[
N_{b}+n_{k}-1\right] };\cdots u^{\mu \left[ n_{k}\right] }\right)
\,,\;k=1,...,I\, \\ 
M_{a_{0}}\left( \cdots q^{b\left[ N_{b}\right] };u^{\mu }\right) =\partial
L/\partial q^{a_{0}}%
\end{array}%
\right. \;.  \label{2.6}
\end{eqnarray}%
Here $M_{a_{k}\,b}\;$is the generalized Hessian matrix and $K_{a_{k}}$ and $%
M_{a_{0}}$ are some LF of the indicated arguments.

Consider the orders of the inner coordinates in the complete set of the ELE.
These orders are $\mathcal{N}_{a}=N_{a}+n_{I}$ . One can see that these
orders are, in fact, defined by a subset of (\ref{2.5}) with $k=I\,.$ In any
subset of the equations (\ref{2.5}) with $k<I\,\,$the orders of the
coordinates are less than in the complete set. Then according to the above
definition, all the ELE with $k<I$ are constraints. The set (\ref{2.5}) has
the following specific structure: In each equation of the complete set the
order of a coordinate $q^{a}$ is the sum of the proper order $N_{a}$ and of
the order $n_{k}$ . The latter is the same for all the coordinates and
depends only on the number $a_{k}$ of the equation.

\subsection{Canonical form}

Let a set of equations%
\begin{equation}
F_{A}\left( \cdots q^{a\left[ \mathcal{N}_{a}\right] }\right) =0\,,
\label{2.8a}
\end{equation}%
be given. Suppose that these equations can be transformed to the following
equivalent form:%
\begin{equation}
q^{\alpha \left[ l_{\alpha }\right] }=\varphi ^{\alpha }\left( \cdots
q^{\alpha \left[ l_{\alpha }-1\right] };\cdots q^{g\left[ l_{g}\right]
}\right) \,,\;\;q^{a}=\left( q^{\alpha },q^{g}\right) \,,\ a=\left( \alpha
,g\right) \,,\;\;l_{a}\leq \mathcal{N}_{a}\;.  \label{2.8b}
\end{equation}%
The equation (\ref{2.8b}) present the canonical form of the initial set (\ref%
{2.8a}).\emph{\ }In the canonical form the equations are solved with respect
to the highest-order time derivatives $q^{\alpha \left[ l_{\alpha }\right] }$
of the coordinates $q^{\alpha }$. The coordinates $q^{g}$ (if they exist)
and their derivatives $q^{g\left[ l_{g}\right] }$ enter into the set (\ref%
{2.8b}) as arbitrary functions of time. In fact, there are no equations for
these coordinates. In what follows we call these coordinates the$\emph{\;}$%
gauge\ coordinates whereas $q^{\alpha }$ we call the\emph{\ }nongauge
coordinates. The orders of the coordinates in the canonical forms may be
less than those in the initial set. In the general case, one and the same
set of equations can have different canonical forms. Generally there are
many canonical form of the same set of equations.

Below, we are going to formulate a general procedure of reducing the ELE to
the canonical form (in what follows it is called the\emph{\ }reduction
procedure). Our consideration is always local in a vicinity of a given
consideration point $q_{0}^{a\left[ l\right] }$ (in the jet space), which is
on shell w.r.t. the ELE. We consider theories and coordinates where the
consideration point could be selected as zero point. Thus, we suppose that
the zero point is on shell. Further we always suppose that the ranks of the
encountered Jacobi matrices\footnote{%
A retangular matrix with elements $\partial A_{\alpha }/\partial x^{i}$ is
often denoted as $\partial A/\partial x\,$\ and called the Jacobi matrix.
\par
{}} are constant in a vicinity of the consideration point. Such suppositions
we call ''suppositions of the ranks''. Saying that some suppositions hold
true in the consideration point, we always suppose that they hold true in a
vicinity of the consideration point. In course of the reduction procedure we
perform several typical transformations with LF or with the corresponding
equations. Each of such transformations lead to equivalent sets of equations
or to equivalent sets of LF (definitions of such equivalences are given
above). The proof of these equivalences is based on two Lemmas which are
presented in the Appendix. Any statement of the form ''the following
equivalence holds true'' can be easily justified by these Lemmas.

\section{Canonical form of nonsingular ELE}

\subsection{A particular case}

Consider theories without external coordinates and with only two different
proper orders of the inner coordinates. In such a case all the indices $a$
can be divided into two groups: $a=\left( a_{1},a_{2}\right) ,$ such that $%
N_{a_{2}}=n_{2}>N_{a_{1}}=n_{1}\,,\;L=L\left( \cdots q^{a_{2}\left[ n_{2}%
\right] },\cdots q^{a_{1}\left[ n_{1}\right] }\right) .$ Consider first the
case $n_{1}>0.$ Then Eqs. (\ref{2.5}) can be written as: 
\begin{eqnarray}
&&F_{a_{2}}=M_{a_{2}\,a}q^{a\left[ N_{a}+n_{2}\right] }+K_{a_{2}}\left(
\cdots q^{b\left[ N_{b}+n_{2}-1\right] }\right) =0\,,  \label{2.15} \\
&&F_{a_{1}}=M_{a_{1}\,a}q^{a\left[ N_{a}+n_{1}\right] }+K_{a_{1}}\left(
\cdots q^{b\left[ N_{b}+n_{1}-1\right] }\right) =0\,.  \label{2.16}
\end{eqnarray}%
The equations (\ref{2.16}) are constraints. Consider the set%
\begin{equation}
M_{a_{1}\,a}q^{a\left[ N_{a}+n_{2}\right] }+K_{a_{1}}^{\left( 1\right)
}\left( \cdots q^{b\left[ N_{b}+n_{2}-1\right] }\right) =0\,,  \label{2.18}
\end{equation}%
obtained from the constraints after being $n_{2}-n_{1}$ times differentiated
with respect to the time $t$. Since $M\neq 0,$ the rectangular matrix $%
M_{a_{1}\,a}$ has a maximal rank, that is why there exists another division
of the indices:%
\begin{equation}
a=\left( a_{|i}\right) ,\;\left[ a_{|i}\right] =\left[ a_{i}\right]
\,,\;i=1,2\,,\;\;\det \,M_{a_{1}b_{|1}}\neq 0\,.  \label{2.18a}
\end{equation}%
Remark that 
\begin{equation}
a_{i}=\left( a_{i|1},a_{i|2}\right) \,,\;a_{|i}=\left(
a_{1|i},a_{2|i}\right) \,,\;\left[ a_{1|2}\right] =\left[ a_{2|1}\right] \,.
\label{2.18b}
\end{equation}%
The set (\ref{2.18}) can be solved with respect to the derivatives $q^{a_{|1}%
\left[ N_{a_{|1}}+n_{2}\right] }$ as follows: 
\begin{equation}
q^{a_{|1}\left[ N_{a_{|1}}+n_{2}\right] }=-\left( M_{1}^{-1}\right)
^{a_{|1}a_{1}}\left[ \left( M_{3}\right) _{a_{1}\,a_{|2}}q^{a_{|2}\left[
N_{a_{|2}}+n_{2}\right] }+K_{a_{1}}^{\left( 1\right) }\left( \cdots q^{b%
\left[ N_{b}+n_{2}-1\right] }\right) \right] .  \label{2.20}
\end{equation}%
Here the matrices $M_{1}$ and $M_{3}$ are defined by the following block
representation of the matrix $M$:%
\begin{equation*}
M_{ab}=\left( 
\begin{array}{cc}
\left( M_{2}\right) _{a_{2}\,b_{|1}} & \left( M_{4}\right) _{a_{2}\,b_{|2}}
\\ 
\left( M_{1}\right) _{a_{1}\,b_{|1}} & \left( M_{3}\right) _{a_{1}\,b_{|2}}%
\end{array}%
\right) \,,\;\;\det \,M_{a_{1}b_{|1}}\neq 0\Longrightarrow \det \,M_{1}\neq
0\,.
\end{equation*}%
Excluding the derivatives $q^{a_{|1}\left[ N_{a_{|1}}+n_{2}\right] }$ from
Eqs. (\ref{2.15}) by the help of (\ref{2.20}), we get the equations%
\begin{equation}
\left( M_{5}\right) _{a_{2}b_{|2}}q^{b_{|2}\left[ N_{b_{|2}}+n_{2}\right]
}+K_{a_{2}}^{\left( 2\right) }\left( \cdots q^{b\left[ N_{b}+n_{2}-1\right]
}\right) =0\,.  \label{2.21}
\end{equation}%
Taking into account an useful relation 
\begin{eqnarray}
\det M &=&\det \left( 
\begin{array}{cc}
M_{2} & M_{4} \\ 
M_{1} & M_{3}%
\end{array}%
\right) =\det \left( 
\begin{array}{cc}
0 & M_{4}-M_{2}M_{1}^{-1}M_{3} \\ 
M_{1} & M_{3}%
\end{array}%
\right)  \label{2.19a} \\
&=&\det M_{1}\det (M_{4}-M_{2}M_{1}^{-1}M_{3})\;,  \notag
\end{eqnarray}%
which is related to the Gaussian reduction of matrices \cite{Gantm59}, we
get:%
\begin{equation}
\left. 
\begin{array}{c}
\det M\neq 0 \\ 
\det M_{1}\neq 0%
\end{array}%
\right\} \Longrightarrow \det \,M_{5}\neq
0,\;\;M_{5}=M_{4}-M_{2}M_{1}^{-1}M_{3}\;.  \label{2.19}
\end{equation}%
Therefore, (\ref{2.21})\ can be solved with respect to the highest-order
derivatives $q^{a_{|2}\left[ N_{a_{|2}}+n_{2}\right] }$ as: 
\begin{eqnarray}
q^{a_{|2}\left[ N_{a_{|2}}+n_{2}\right] } &=&-\left( M_{5}^{-1}\right)
^{a_{|2}a_{2}}\left[ K_{a_{2}}\left( \cdots q^{b\left[ N_{b}+n_{2}-1\right]
}\right) -\left( M_{2}\;M_{1}^{-1}\right)
_{a_{2}}^{a_{1}}K_{a_{1}}^{(1)}\left( \cdots q^{b\left[ N_{b}+n_{2}-1\right]
}\right) \right]  \notag \\
&\equiv &\varphi ^{a_{|2}}\left( \cdots q^{b_{|2}\left[ N_{b_{|2}}+n_{2}-1%
\right] },\cdots q^{b_{|1}\left[ N_{b_{|1}}+n_{2}-1\right] }\right) \,.
\label{2.22}
\end{eqnarray}%
Thus, we get a set%
\begin{eqnarray}
&&\psi ^{a_{|2}}=q^{a_{|2}\left[ N_{a_{|2}}+n_{2}\right] }-\varphi
^{a_{|2}}\left( \cdots q^{b_{|2}\left[ N_{b_{|2}}+n_{2}-1\right] },\cdots
q^{b_{|1}\left[ N_{b_{|1}}+n_{2}-1\right] }\right) =0\,,  \label{2.23} \\
&&F_{a_{1}}=\left( M_{1}\right) _{a_{1}\,a_{|1}}q^{a_{|1}\left[
N_{a_{|1}}+n_{1}\right] }-K_{a_{1}}^{\left( 3\right) }\left( \cdots q^{b_{|2}%
\left[ N_{b_{|2}}+n_{1}\right] },\cdots q^{b_{|1}\left[ N_{b_{|1}}+n_{1}-1%
\right] }\right) =0\,,  \label{2.24}
\end{eqnarray}%
which is strong equivalent to the initial ELE by virtue of the Lemma 1.

Due to the condition $\det M_{1}\neq 0,$ the equations (\ref{2.24}) can be
solved with respect to $q^{a_{|1}\left[ N_{a_{|1}}+n_{1}\right] }$ and we
obtain: 
\begin{eqnarray}
q^{a_{|1}\left[ N_{a_{|1}}+n_{1}\right] } &=&-\left( M_{1}^{-1}\right)
^{a_{|1}a_{1}}\left[ \left( M_{3}\right) _{a_{1}\,a_{|2}}q^{a_{|2}\left[
N_{a_{|2}}+n_{1}\right] }+K_{a_{1}}\left( \cdots q^{b\left[ N_{b}+n_{2}-1%
\right] }\right) \right]  \notag \\
&\equiv &f^{a_{|1}}\left( \cdots q^{b_{|2}\left[ N_{b_{|2}}+n_{1}\right]
},\cdots q^{b_{|1}\left[ N_{b_{|1}}+n_{1}-1\right] }\right) \,.  \label{2.26}
\end{eqnarray}%
Eqs. (\ref{2.23}) and (\ref{2.26}) are not of canonical form since the
functions $\varphi ^{a_{|2}}$ contain derivatives $q^{b_{|1}\left[
N_{b_{|1}}+n_{2}-1\right] }$ exceeding the ''allowed'' order $\left[
N_{b_{|1}}+n_{1}-1\right] $. Now we exclude all the surplus derivatives $%
q^{a_{|1}\left[ N_{a_{|1}}+n_{1}\right] },...,q^{a_{|1}\left[
N_{a_{|1}}+n_{2}-1\right] }$ from the right hand side of (\ref{2.23}) by the
help of (\ref{2.26}) and corresponding derivatives of it. To this end we
need to differentiate (\ref{2.26}) not more than $n_{2}-n_{1}-1$ times.
Finally, we obtain the following strong equivalent form (the equivalence is
justified by the Lemma 1) of the ELE: 
\begin{eqnarray}
&&q^{a_{|2}\left[ N_{a_{|2}}+n_{2}\right] }=f^{a_{|2}}\left( \cdots q^{b_{|2}%
\left[ N_{b_{|2}}+n_{2}-1\right] },\cdots q^{b_{|1}\left[ N_{b_{|1}}+n_{1}-1%
\right] }\right) ,  \notag \\
&&q^{a_{|1}\left[ N_{a_{|1}}+n_{1}\right] }=f^{a_{|1}}\left( \cdots q^{b_{|2}%
\left[ N_{b_{|2}}+n_{1}\right] },\cdots q^{b_{|1}\left[ N_{b_{|1}}+n_{1}-1%
\right] }\right) \,.  \label{2.27}
\end{eqnarray}%
It is just the canonical form. Taking into account the division of the
indices w.r.t. proper orders of the coordinates, one gets: 
\begin{eqnarray}
&&q^{a_{2|2}\left[ 2n_{2}\right] }=f^{a_{2|2}}\left( \cdots q^{b_{2|2}\left[
2n_{2}-1\right] },\cdots q^{b_{1|2}\left[ n_{1}+n_{2}-1\right] },\cdots
q^{b_{2|2}\left[ n_{2}+n_{1}-1\right] },\cdots q^{b_{1|1}\left[ 2n_{1}-1%
\right] }\right) \,,  \notag \\
&&q^{a_{1|2}\left[ n_{1}+n_{2}\right] }=f^{a_{1|2}}\left( \cdots q^{b_{2|2}%
\left[ 2n_{2}-1\right] },\cdots q^{b_{1|2}\left[ n_{1}+n_{2}-1\right]
},\cdots q^{b_{2|1}\left[ n_{2}+n_{1}-1\right] },\cdots q^{b_{1|1}\left[
2n_{1}-1\right] }\right) \,,  \notag \\
&&q^{a_{2|1}\left[ n_{2}+n_{1}\right] }=f^{a_{2|1}}\left( \cdots q^{b_{2|2}%
\left[ n_{1}+n_{2}\right] },\cdots q^{b_{1|2}\left[ 2n_{1}\right] },\cdots
q^{b_{2|1}\left[ n_{2}+n_{1}-1\right] },\cdots q^{b_{1|1}\left[ 2n_{1}-1%
\right] }\right) \,,  \notag \\
&&q^{a_{1|1}\left[ 2n_{1}\right] }=f^{a_{1|1}}\left( \cdots q^{b_{2|2}\left[
n_{1}+n_{2}\right] },\cdots q^{b_{1|2}\left[ 2n_{1}\right] },\cdots
q^{b_{2|1}\left[ n_{2}+n_{1}-1\right] },\cdots q^{b_{1|1}\left[ 2n_{1}-1%
\right] }\right) \,.  \label{2.27a}
\end{eqnarray}

Remark that the number of the initial data is equal to $2\sum_{a}N_{a}\,.$
Indeed,%
\begin{eqnarray*}
&&\left[ a_{2|2}\right] \left( n_{2}+n_{2}\right) +\left[ a_{1|2}\right]
\left( n_{1}+n_{2}\right) +\left[ a_{2|1}\right] \left( n_{2}+n_{1}\right) +%
\left[ a_{1|1}\right] \left( n_{1}+n_{1}\right) \\
&&\,=2\left[ a_{2}\right] n_{2}+2\left[ a_{1}\right] n_{1}=2\sum_{a}N_{a}\,.
\end{eqnarray*}

One ought to mention that the canonical form (\ref{2.27a}) was obtained in %
\cite{GitTyL85}. However, the procedure that was used for that purpose did
not provide the proof of the equivalence between the initial ELE and the
form (\ref{2.27a}).

Suppose now that the Lagrange function contains degenerate coordinates $%
q^{a_{0}}$,\ $a=\left( a_{0},a_{1}\right) .$ Thus, $L=L\left(
q^{a_{0}},\cdots q^{a_{_{1}}\left[ n_{1}\right] }\right) $ and the ELE read: 
\begin{eqnarray}
&&F_{a_{1}}\equiv M_{a_{1}\,a}q^{a\left[ N_{a}+n_{1}\right]
}+K_{a_{1}}\left( \cdots q^{b\left[ N_{b}+n_{1}-1\right] }\right) =0\,,
\label{2.15c} \\
&&F_{a_{0}}\equiv \frac{\partial L}{\partial q^{a_{0}}}=M_{a_{0}}\left(
\cdots q^{b\left[ N_{b}\right] }\right) =0\,.  \label{2.16c}
\end{eqnarray}%
Despite these equations are formally different from the above case, the
whole procedure of reductions goes through without any essential change. In
fact, differentiating Eq. (\ref{2.16c}) $n_{1}$ times, one obtains 
\begin{equation}
M_{a_{0}\,a}q^{a\left[ N_{a}+n_{1}\right] }+K_{a_{0}}^{(1)}\left( \cdots q^{b%
\left[ N_{b}+n_{1}-1\right] }\right) =0,  \label{2.18c}
\end{equation}%
and all the previous steps may be done as before. Namely, one obtains 
\begin{equation}
q^{a_{|1}\left[ N_{a_{|1}}+n_{1}\right] }=\varphi ^{a_{|1}}\left( \cdots
q^{b_{|1}\left[ N_{b_{|1}}+n_{1}-1\right] },\cdots q^{b_{|0}\left[
N_{b_{|0}}+n_{0}-1\right] }\right) \,,  \label{2.22d}
\end{equation}%
and, since $\det ||(M_{1})_{a_{0}\,a_{|0}}||\neq 0$, Eqs.~(\ref{2.16c}) can
be solved with respect to the variable $q^{a_{|0}\left[ N_{a_{|0}}\right] }$
as follows: 
\begin{equation*}
q^{a_{|0}\left[ N_{a_{|0}}\right] }=f^{a_{|0}}\left( \cdots q^{b_{|1}\left[
N_{b_{|1}}\right] },\cdots q^{b_{|0}\left[ N_{b_{|0}}-1\right] }\right) \,.
\end{equation*}%
Finally, after eliminating the ``bad'' derivatives in the right hand side of
(\ref{2.22d}) for $q^{a_{|1}\left[ N_{a_{|1}}+n_{1}\right] }$ one ends up
again with Eqs.~(\ref{2.27a}) but now with $n_{2}\rightarrow
n_{1},\;n_{1}\rightarrow 0$ (by convention: $q^{b_{1|1}\left[ -1\right]
}\equiv 0$).

\subsection{General nonsingular ELE}

Consider the general nonsingular ELE. Here the Lagrange function may contain
some degenerate inner coordinates, higher derivatives of some inner
coordinates, and, moreover, may depend on some external coordinates, $%
L=L\left( \cdots q^{a\left[ N_{a}\right] };u^{\mu }\right) \,,\;N_{a}\geq
0\,.$ Thus, we are going to deal with the nonsingular ELE of the form (\ref%
{2.5}). Our aim is to present these equations in an equivalent canonical
form.

{\LARGE Theorem\ 1: }{\large The nonsingular ELE}\emph{\ }{\large (\ref{2.5}%
) can be transformed to the following equivalent canonical form:}%
\begin{eqnarray}
&&f^{a_{i|k}}=q^{a_{i|k}\left[ n_{i}+n_{k}\right] }-\varphi ^{a_{i|k}}\left(
\cdots q^{b_{j|k_{-}}\left[ n_{j}+n_{k_{-}}-1\right] },\cdots q^{b_{j|k_{+}}%
\left[ n_{j}+n_{k}\right] };\cdots u^{\mu \left[ n_{k}\right] }\right) =0\,,
\notag \\
&&I\geq k_{+}\geq k+1,\,\;k\geq k_{-}\geq 0,\;i,j,k=0,1,...,I\,,
\label{2.28b}
\end{eqnarray}%
{\large where the indices\ of the coordinates are divided into groups as
follows: }$a=\left( a_{i}\right) ${\large \ is the division of the indices
w.r.t. the proper orders of the coordinates, and besides} 
\begin{equation*}
a_{i}=\left( a_{i|k}\,,\;i,k=0,1,...,I\,\right) \,,\;\;\left[ a_{i|k}\right]
\geq 0,\;\;\sum_{k}\left[ a_{i|k}\right] =\sum_{k}\left[ a_{k|i}\right] =%
\left[ a_{i}\right] =\left[ a_{|i}\right] \,.
\end{equation*}%
{\large Moreover, the equivalence }$F\sim f$ \ {\large between the
corresponding LF holds true.} {\large That implies}%
\begin{equation*}
F_{a}=\hat{U}_{ab}f^{b}\,,\;f^{b}=\hat{V}^{ba}F_{a}\,,\;\;\hat{U}_{ab}\hat{V}%
^{bc}=\delta _{a}^{c}\,,
\end{equation*}%
$\;$ {\large where} $\hat{U}$ {\large and} $\hat{V}$ {\large are LO}.
Besides, that implies the strong equivalence between the ELE and their
canonical form (\ref{2.28b}).

The proof of the Theorem 1 may be considered, in fact, as the general
reduction procedure to the canonical form for the nonsingular ELE.

It is reasonable to divide the reduction procedure into two parts. These
parts may be called conditionally ''the preliminary resolution'', and ''the
subordination procedure''.%
\begin{equation*}
{\Large Preliminary\ resolution}
\end{equation*}

Let us introduce the notation $a=\left( \underline{a},a_{I}\right) \,,\;%
\underline{a}=\left( a_{k}\,,\;k=0,1,...,I\,-1\right) \,,\;N_{\underline{a}%
}<n_{I}\,,$ such that the ELE read: 
\begin{eqnarray}
\hspace{-1cm} &&F_{a_{I}}\left( \cdots q^{b\left[ N_{b}+n_{I}\right]
};\cdots u^{\mu \left[ n_{I}\right] }\right) =M_{a_{I}\,b}q^{b\left[
N_{b}+n_{I}\right] }+K_{a_{I}}\left( \cdots q^{b\left[ N_{b}+n_{I}-1\right]
};\cdots u^{\mu \left[ n_{I}\right] }\right) =0\,,  \label{2.28} \\
\hspace{-1cm} &&F_{\underline{a}}\left( \cdots q^{b\left[ N_{b}+N_{%
\underline{a}}\right] };\cdots u^{\mu \left[ N_{\underline{a}}\right]
}\right) =0\,.  \label{2.29c}
\end{eqnarray}%
Recall that the equations (\ref{2.29c}) can be considered as constraints.

The first step of the procedure is the following: We consider the
consistency conditions of the constraints. Namely, we consider the equations
that are obtained from the constraints by differentiating them $n_{I}-n_{%
\underline{a}}$ times, 
\begin{equation}
F_{\underline{a}}^{\left[ n_{I}-N_{\underline{a}}\right] }=M_{\underline{a}%
\,b}q^{b\left[ N_{b}+n_{I}\right] }+K_{\underline{a}}^{\left( 1\right)
}\left( \cdots q^{b\left[ N_{b}+n_{I}-1\right] };\cdots u^{\mu \left[ n_{I}%
\right] }\right) =0\,.  \label{2.30}
\end{equation}%
Here $K_{\underline{a}}^{\left( 1\right) }$ are some LF of the indicated
arguments. Remark that the orders of all the coordinates in the set (\ref%
{2.30}) coincide with the ones in the complete set. For $M\neq 0$, the
matrix 
\begin{equation}
\frac{\partial F_{a}^{\left[ n_{I}-N_{a}\right] }}{\partial q^{b\left[
N_{b}+n_{I}\right] }}=\frac{\partial ^{2}L}{\partial q^{a\left[ N_{a}\right]
}\partial q^{b\left[ N_{b}\right] }}=M_{a\,b}  \label{2.29a}
\end{equation}%
is invertible. At the same time, the rectangular matrix $M_{\underline{a}%
\,a} $ has the maximal rank $\left[ \underline{a}\right] $. Therefore, there
exists a division of the indices $a$ such that: 
\begin{equation}
a=\left( \bar{a},a_{|I}\right) ,\;\left[ \bar{a}\right] =\left[ \underline{a}%
\right] ,\;\left[ a_{|I}\right] =\left[ a_{I}\right] \,,\;\det \,M_{%
\underline{a}\bar{a}}\neq 0\,.  \label{2.32a}
\end{equation}%
Thus, the division (\ref{2.3a}) of the indices $a$ w.r.t. the coordinate
proper orders becomes more detailed,%
\begin{eqnarray*}
&&a_{i}=\left( \bar{a}_{i},a_{i|I}\right) \,,\;\bar{a}=\left( \bar{a}%
_{i}\right) ,\;a_{|I}=\left( a_{i|I}\right) ,\;\left[ a_{i|I}\right] \geq
0\,, \\
&&\sum_{i}\left[ a_{i|I}\right] =\left[ a_{|I}\right] =\left[ a_{I}\right]
,\;\sum_{i}\left[ \bar{a}_{i}\right] =\left[ \bar{a}\right] =\left[ 
\underline{a}\right] \,.
\end{eqnarray*}

Due to (\ref{2.32a}), the set (\ref{2.30}) can be solved with respect to the
derivatives $q^{\bar{a}\left[ N_{\bar{a}}+n_{I}\right] }$ as: 
\begin{equation}
q^{\bar{a}\left[ N_{\bar{a}}+n_{I}\right] }=-\left( M_{1}^{-1}\right) ^{%
\bar{a}\underline{a}}\left[ \left( M_{3}\right) _{\underline{a}%
\,b_{|I}}q^{b_{|I}\,\left[ N_{b_{|I}}+n_{I}\right] }+K_{\underline{a}%
}^{\left( 1\right) }\left( \cdots q^{b\left[ N_{b}+n_{I}-1\right] };\cdots
u^{\mu \left[ n_{I}\right] }\right) \right] \,,  \label{2.32}
\end{equation}%
where 
\begin{equation*}
M_{ab}=\left( 
\begin{array}{cc}
\left( M_{2}\right) _{a_{I}\,\bar{b}} & \left( M_{4}\right) _{a_{I}\,b_{|I}}
\\ 
\left( M_{1}\right) _{\underline{a}\,\bar{b}} & \left( M_{3}\right) _{%
\underline{a}\,b_{|I}}%
\end{array}%
\right) \;.
\end{equation*}

Excluding the derivatives $q^{\bar{a}\left[ N_{\bar{a}}+n_{I}\right] }$ from
Eqs. (\ref{2.28}) by the help of (\ref{2.32}), we get the set: 
\begin{eqnarray}
&&\left( M_{5}\right) _{a_{I}b_{|I}}q^{b_{|I}\left[ N_{b_{|I}}+n_{I}\right]
}+K_{a_{I}}^{\left( 2\right) }\left( \cdots q^{b\left[ N_{b}+n_{I}-1\right]
};\cdots u^{\mu \left[ n_{I}\right] }\right) =0\,,  \notag \\
&&M_{5}=M_{4}-M_{2}M_{1}^{-1}M_{3}\,,\;\det M_{5}\neq 0\,,  \label{2.34}
\end{eqnarray}%
where $K_{a_{I}}^{\left( 2\right) }$ are some LF of the indicated arguments.
The set (\ref{2.34}) can be solved with respect to its highest-order
derivatives $q^{a_{|I}\left[ N_{a_{|I}}+n_{I}\right] }$ as: 
\begin{equation}
q^{a_{|I}\left[ N_{a_{|I}}+n_{I}\right] }=\phi ^{a_{|I}}\left( \cdots
q^{b_{|I}\left[ N_{b_{|I}}+n_{I}-1\right] },\cdots q^{\bar{b}\left[ N_{\bar{b%
}}+n_{I}-1\right] };\cdots u^{\mu \left[ n_{I}\right] }\right) \,,
\label{2.33}
\end{equation}%
where $\varphi ^{a_{|I}}$ are some LF. Thus, after the first step we get a
set of equations 
\begin{eqnarray}
&&\psi ^{a_{|I}}=q^{a_{|I}\left[ N_{a_{|I}}+n_{I}\right] }-\phi
^{a_{|I}}\left( \cdots q^{b_{|I}\left[ N_{b_{|I}}+n_{I}-1\right] },\cdots q^{%
\bar{b}\left[ N_{\bar{b}}+n_{I}-1\right] };\cdots u^{\mu \left[ n_{I}\right]
}\right) =0\,,  \label{2.35} \\
&&F_{\underline{a}}\left( \cdots q^{b\left[ N_{b}+N_{\underline{a}}\right]
};\cdots u^{\mu \left[ N_{\underline{a}}\right] }\right) =0\,,\;\underline{a}%
=\left( a_{k}\,,\;k=0,1,...,I\,-1\right) \,,\;N_{\underline{a}}<N_{I}\,\,,
\label{2.36}
\end{eqnarray}%
\ which are strong equivalent to the initial ELE by virtue of the Lemma 1
from the Appendix.

At the second step we turn to the subset (\ref{2.36}). We remark that this
subset has the same structure as the complete initial set of the ELE if one
considers the coordinates $q^{\bar{a}}$ as inner ones and the variables $%
q^{a_{|I}}\,$as external ones. Namely, let us denote 
\begin{eqnarray*}
&&\overset{1}{F}_{\underline{a}}\left( \cdots q^{\bar{b}\left[ N_{\bar{b}%
}+N_{\underline{a}}\right] };\cdots u^{\mu _{1}\left[ N_{\underline{a}}%
\right] }\right) =F_{\underline{a}}\left( \cdots q^{b\left[ N_{b}+N_{%
\underline{a}}\right] };\cdots u^{\mu \left[ N_{\underline{a}}\right]
}\right) \,, \\
&&u^{\mu _{1}}=\left( u^{\mu },\cdots q^{a_{|I}\left[ N_{a_{|I}}\right]
}\right) ,\;\mu _{1}=\left( \mu ,a_{|I}\right) \,.
\end{eqnarray*}%
Then the set (\ref{2.36}) can be written as:%
\begin{equation}
\overset{1}{F}_{\underline{a}}\left( \cdots q^{\bar{b}\left[ N_{\bar{b}}+N_{%
\underline{a}}\right] };\cdots u^{\mu _{1}\left[ N_{\underline{a}}\right]
}\right) =0\,,\;\underline{a}=\left( a_{k}\,,\;k=0,1,...,I\,-1\right)
\,,\;N_{\underline{a}}<N_{I}\,\,,  \label{2.36a}
\end{equation}%
where%
\begin{equation*}
\overset{1}{F}_{a_{k}}=\left\{ 
\begin{array}{c}
M_{a_{k}\,\bar{b}}q^{\bar{b}\left[ N_{\bar{b}}+n_{k}\right] }+\underline{K}%
_{a_{k}}\left( \cdots q^{\bar{b}\left[ N_{\bar{b}}+n_{k}-1\right] };\cdots
u^{\mu _{1}\left[ n_{k}\right] }\right) ,\;k=1,...,I\,-1 \\ 
\underline{M}_{a_{0}}\left( \cdots q^{\bar{b}\left[ N_{\bar{b}}\right]
};u^{\mu _{1}}\right) =\partial L/\partial q^{a_{0}}%
\end{array}%
\right. \;.
\end{equation*}%
Here $q^{\bar{a}}$ are the inner coordinates, and $u^{\mu _{1}}$ are the
external coordinates. The order of the set (\ref{2.36a}) is $2n_{I-1}.$
Furthermore, by virtue of (\ref{2.32a}), the matrix%
\begin{equation}
\frac{\partial \overset{1}{F}_{\underline{a}}^{\left[ n_{I-1}-N_{\underline{a%
}}\right] }}{\partial q^{\bar{b}\left[ N_{\bar{b}}+n_{I-1}\right] }}=M_{%
\underline{a}\,\bar{b}}  \label{2.37}
\end{equation}%
is invertible. Thus, the structure (\ref{2.5},\ref{2.6}) is repeated
completely.

At the same time, the number of the inner variables, the number of the
equations, and the order of the set (\ref{2.36a}) are less than those of the
initial set of the ELE (\ref{2.5},\ref{2.6}).

Now, we apply the same procedure as in the first step to the reduced set (%
\ref{2.36a}). That will be the second step of the reduction procedure. It
will produce equations of similar structure with less inner variables and of
lower order. After the last $(I+1)$-th step the ELE (\ref{2.5}) may be
written in the following strong equivalent form:%
\begin{eqnarray}
&&q^{a_{i|k}\left[ n_{i}+n_{k}\right] }=\phi ^{a_{i|k}}\left( \cdots
q^{b_{j|k_{+}}\left[ n_{j}+n_{k}\right] },\cdots q^{b_{j|k_{-}}\left[
n_{j}+n_{k}-1\right] };\cdots u^{\mu \left[ n_{k}\right] },\right) ,  \notag
\\
&&I\geq k_{+}\geq k+1\,,\;k\geq k_{-}\geq 0\,\,,I\geq i,j\geq 0\,,
\label{2.39}
\end{eqnarray}%
where $\phi ^{a_{i|k}}$ are some LF of the indicated arguments (the
arguments $\cdots q^{b_{j|k_{+}}\left[ n_{j}+n_{k}\right] }$ result from
those coordinates that intermediately have been considered as external
ones), and the indices $a_{i}$ are divided into the following groups:%
\begin{equation*}
a_{i}=\left( a_{i|k}\right) \,,\;\;\left[ a_{i|k}\right] \geq 0,\;\;\sum_{k}%
\left[ a_{i|k}\right] =\sum_{k}\left[ a_{k|i}\right] =\left[ a_{i}\right]
\,,\;\;i,k=0,1,...,I\,.
\end{equation*}

The set (\ref{2.39}) is still not the canonical form of the ELE. The reason
is that the right hand sides of the set contain derivatives of orders that
may exceed the orders $n_{i}+n_{k}$ of the (highest) derivatives $q^{a_{i|k}%
\left[ n_{i}+n_{k}\right] }$ appearing on the left hand side of the set. We
recall that by the definition in the canonical form there is a subordination
of derivative orders, namely, the orders of all the derivatives in the right
hand sides have to be less than the ones on the left hand side. Explicitly,
this subordination would require that the following inequalities should
hold: 
\begin{eqnarray}
&&n_{j}+n_{k_{+}}>n_{j}+n_{k}\,,  \notag \\
&&n_{j}+n_{k_{-}}>n_{j}+n_{k}-1\,,  \notag
\end{eqnarray}%
which, because of the inequalities $n_{I}>n_{I-1}>\cdots n_{1}>n_{0}$ , is
true for the first line and the case $k_{-}=k$ of the second line, and it is
definitely not true for the cases $k_{-}<k$. Arranging the equations (\ref%
{2.39}) (for fixed value of $i$) in descending order w.r.t. $k$, and the
arguments in the functions $\varphi $ (for fixed value of $j$) also in
descending order w.r.t. the value of $k_{+}$ and $k_{-}$, we get, when
disregarding the common value $n_{j}$, a quadratic matrix whose main
diagonal (i.e. elements with $k=k_{-}$) contains the entries $n_{k}-1$,
whereas the entries to the left of that diagonal are equal to $n_{k}$, and
to the right of that diagonal are equal to $n_{k}-1$. Therefore, below the
main diagonal occur ``good'' derivatives, and above it occur ``bad''
derivatives not obeying the subordination requirement.%
\begin{equation*}
{\Large Subordination\ procedure}
\end{equation*}

One can see that these ''bad'' derivatives can be excluded from the right
hand sides by the help of corresponding ''lower'' equations of the set and
their differential consequences (compare Eqs. (\ref{2.23}) and (\ref{2.26})
for the simple case $I=2$). In what follows we call such an exclusion the
subordination procedure.

In order to be more definite let us write down two arbitrary lines, $\ell >k$%
, of the right hand sides of the set of equations (\ref{2.39}) (for the
highest derivatives only):%
\begin{eqnarray*}
\hspace{-1cm} &&\phi ^{a_{i|\ell }}\left( q^{b_{j|I}\left[ n_{j}+n_{\ell }%
\right] },...,q^{b_{j|\ell +1}\left[ n_{j}+n_{\ell }\right] },q^{b_{j|\ell }%
\left[ n_{j}+n_{\ell }-1\right] },...,q^{b_{j|k+1}\left[ n_{j}+n_{\ell }-1%
\right] },q^{b_{j|k}\left[ n_{j}+n_{\ell }-1\right] },...,q^{b_{j|0}\left[
n_{j}+n_{\ell }-1\right] }\right) , \\
\hspace{-1cm} &&\vdots \\
\hspace{-1cm} &&\phi ^{a_{i|k}}\left( q^{b_{j|I}\left[ n_{j}+n_{k}\right]
},...,q^{b_{j|\ell +1}\left[ n_{j}+n_{k}\right] },\;\;q^{b_{j|\ell }\left[
n_{j}+n_{k}\right] },...,q^{b_{j|k+1}\left[ n_{j}+n_{k}\right]
},\;\;q^{b_{j|k}\left[ n_{j}+n_{k}-1\right] },...,q^{b_{j|0}\left[
n_{j}+n_{k}-1\right] }\right) .
\end{eqnarray*}

Obviously, because $n_{\ell }>n_{k}$ all the derivatives of the equation for 
$q^{a_{i|\ell }\left[ n_{i}+n_{\ell }\right] }$ with $k\geq \ell _{-}\geq 0$
are ``bad'' with respect to the derivatives $q^{a_{i|k}\left[ n_{i}+n_{k}%
\right] }$ (remind $\ell \geq \ell _{-}\geq 0$). However, these ``bad''
derivatives can be eliminated by the equations for the latter ones, $%
q^{a_{i|k}\left[ n_{i}+n_{k}\right] }$, and their differential consequences
up to the order $n_{\ell }-n_{k}-1$. Thereby, the function $\phi ^{a_{i|\ell
}}$ changes into some new function $\tilde{\phi}^{a_{i|\ell }}$. One can see
that doing this we do not change the highest orders of derivatives of the
other coordinates, both proper and external ones, in the right hand side of
the equation for $q^{a_{i|\ell }\left[ n_{i}+n_{\ell }\right] }$. (Remind,
that the derivatives of the external coordinates are $u^{\mu \left[ n_{\ell }%
\right] }$ and $u^{\mu \left[ n_{k}\right] }$, respectively.)

This subordination procedure, starting with $\ell = I$ may be done for any $%
k < I$, thereby ``cleaning'' every entry on the right hand side of equations
for $q^{ a_{i|I}\left[n_{i}+n_{I}\right] }$. Namely, the highest orders of
derivatives on the r.h.s. become $q^{ b_{j|k_-}\left[n_{i}+n_{k_-} - 1\right]
}$ with $I \geq k_- \geq 0$ (for the case $\ell = I$ no $k_+$ appears). Then
the procedure will be applied to the equations for $q^{ a_{i|I-1}\left[%
n_{i}+n_{I-1}\right] }$, and so forth, up to $q^{ a_{i|0}\left[n_{i}+n_{0}%
\right] }$, where nothing is to be changed.

After having eliminated all the ``bad'' derivatives, we transformed the set (%
\ref{2.39}), and therefore the initial ELE, to the following strong
equivalent (the equivalence is justified by the Lemma 1) canonical form%
\begin{eqnarray*}
&&q^{a_{i|k}\left[ n_{i}+n_{k}\right] }=\varphi ^{a_{i|k}}\left( \cdots
q^{b_{j|k_{+}}\left[ n_{j}+n_{k}\right] },\cdots q^{b_{j|k_{-}}\left[
n_{j}+n_{k_{-}}-1\right] };\cdots u^{\mu \left[ n_{k}\right] }\right) \,, \\
&&I\geq k_{+}\geq k+1,\;k\geq k_{-}\geq 0,\;i,j,k=0,1,...,I\,,
\end{eqnarray*}%
where $\varphi ^{a_{i|k}}$ are some LF of the indicated arguments. This
proves the Theorem 1.

We see that there are no gauge coordinates in the nonsingular ELE.

The number of the initial data is equal to $2\sum_{a}N_{a}\,.$ Indeed,%
\begin{equation*}
\sum_{i,k}\left[ a_{i|k}\right] \left( N_{i}+N_{k}\right) =\sum_{i}\left(
N_{i}\sum_{k}\left[ a_{i|k}\right] \right) +\sum_{k}\left( N_{k}\sum_{i}%
\left[ a_{i|k}\right] \right) =2\sum_{a}N_{a}.
\end{equation*}

One ought to remark that in the general case there exist many different
canonical forms of the nonsingular ELE. This uncertainty is related to the
possibility of different choices of nonzero minors of a matrix with a given
rank (different divisions of the indices $a_{i}$ \ in course of the
reduction procedure). However, as it was demonstrated above, the number of
the equations in the canonical form (which is equal to the number of the ELE
in the nonsingular case) and the number of the initial data is the same for
all possible canonical forms.

\section{Canonical form of singular ELE}

Studying the canonical form of nonsingular ELE, we have demonstrated that
the equations in the canonical form are solved with respect to the
highest-order derivatives $q^{a_{i|k}\left[ n_{i}+n_{k}\right] },$ where $%
n_{i}$ are the proper orders of the coordinates $q^{a_{i}}$ . However,
considering specific examples, one can see that this is not always true for
singular ELE. Namely, in the canonical form of the latter case, the highest
orders of the derivatives $q^{a_{i}\left[ l\right] }$ may take on all the
values from zero to $n_{i}+I$ . The reduction procedure to the canonical
form for the general singular ELE is considered below. In the singular case,
already after the first step of the reduction procedure, the ELE cease to
have their initial specific structure (\ref{2.5},\ref{2.6}). Namely, the
simple structure of terms with highest-order derivatives in the equations
may be lost. That is why in the singular case it is more convenient to
formulate the reduction procedure for a more general set of ordinary
differential equations, which contains the ELE as a particular case. Namely,
further we are going to consider a set of the form\footnote{%
We do not indicate here possible external coordinates.}: 
\begin{equation}
F_{A_{\mu }}\left( \cdots q^{a_{i}\left[ i+\mu \right] }\right)
=0\,;\;\;i=0,1,...,I\,,\;\mu =0,...,J\,.  \label{3.1}
\end{equation}%
Here $F_{A_{\mu }}\left( \cdots q^{a_{i}\left[ i+\mu \right] }\right) $ are
some LF. Via $a_{i}$ and $A_{\mu }$ are denoted sets of indices, $\left[
a_{i}\right] \geq 0,\;\left[ A_{\mu }\right] \geq 0\,$, and the complete set
of the inner coordinates in Eqs. (\ref{3.1}) is $q^{a}=\left(
q^{a_{0}},...,q^{a_{I}}\right) \,,$\ $a=\left(
a_{i}\,,\;\;i=0,1,...,I\right) .$ The indices $A=\left( A_{\mu }\right) $
enumerate the equations. In the general case the number of the indices $A$
(the number of all the equations) is not equal to the number of the indices $%
a$ (the number of the coordinates).{\LARGE \ }The division of the indices $A$
into the groups is not related to the division of the indices $a$ into the
groups. The orders of the coordinates $q^{a_{i}}$ in the complete set (\ref%
{3.1}) are: $\mathcal{N}_{a_{i}}=i+J\,.$ In fact, these orders are defined
by a subset of (\ref{3.1}) with $\mu =J.$ In all the other equations with $%
\mu <J\,\ $the coordinates $q^{a_{i}}$ have the orders less than $i+J$.
Thus, the latter equations are constraints.

Similar to the ELE (\ref{2.5}), the set (\ref{3.1}) has the following
specific structure: In each equation of the set the order of a coordinate $%
q^{a_{i}}$ is the sum of the proper order $i$ and of the order $\mu $. The
latter is the same for all the coordinates and is related to the number of
the equation in the set.

Below we consider the reduction procedure to the canonical form for the
equations (\ref{3.1}). In fact, this reduction procedure is formulated in
the proof of the Theorem 2 given below. The Theorem 2 holds true under
certain suppositions of the structure of the functions $F_{A_{\mu }}\,.$
These suppositions are formulated as suppositions of the ranks of some
Jacobi matrices involving the functions $F_{A_{\mu }}\,.$ First of all, the
complete matrix%
\begin{equation}
M_{A_{\mu }\,a_{i}}=\frac{\partial F_{A_{\mu }}}{\partial q^{a_{i}\left[
i+\mu \right] }}=\frac{\partial F_{A_{\mu }}^{\left[ J-\mu \right] }}{%
\partial q^{a_{i}\left[ i+J\right] }}\,,  \label{3.2}
\end{equation}%
has to have a constant rank in a vicinity of the consideration point (one
can see that the matrix $M_{A_{\mu }\,a_{i}}$ coincides with generalized
Hessian matrix if the set (\ref{3.1}) is the Lagrangian one).

{\LARGE Theorem 2}: {\large Under certain suppositions of the ranks, the
equations (\ref{3.1}) can be transformed to the following equivalent
canonical form:}%
\begin{eqnarray}
&&f^{a_{i|\sigma }}=q^{a_{i|\sigma }\left[ i+\sigma \right] }-\varphi
^{a_{i|\sigma }}\left( \cdots q^{a_{j|\sigma _{-}}\left[ j+\sigma _{-}-1%
\right] },\cdots q^{a_{j|\sigma _{+}}\left[ j+\sigma \right] }\right) =0\,, 
\notag \\
&&i,j=0,1,...,I\,,\;\;\sigma =-I,...,J\,,\;\;-I\leq \sigma _{-}\leq \sigma
\,,\;\sigma +1\leq \sigma _{+}\leq J+1\,,\;  \label{3.3}
\end{eqnarray}%
{\large where all the indices }$a${\large \ are divided into groups as
follows:} 
\begin{equation}
a_{i}=\left( a_{i|\sigma }\right) \,,\;\;\left[ a_{i|\sigma }\right] \geq
0\,,\;\;\sigma =-I,...,J+1,\;\;\left( \left[ a_{i|\sigma }\right] =0\;\;%
\mathrm{if}\;\;i+\sigma <0\right) \,,  \label{3.4}
\end{equation}%
{\large and it is thought that negative powers of the time derivatives do
not exist, that is: }$\left[ q^{a\left[ p\right] }\right] =0${\large \ for} $%
p<0${\large .}

{\large Moreover, the following equivalence between the corresponding LF
holds true:}%
\begin{eqnarray}
&&F_{A}\sim \bar{F}_{A}=\left( 
\begin{array}{c}
f^{a_{i|\sigma }} \\ 
0_{G}%
\end{array}%
\right) \,,\;\;A=\left( a_{i|\sigma }\,,G\right) \,,\;i=0,1,...,I\,,\;\sigma
=-I,...,J\,,  \notag \\
&&0_{G}\equiv 0\,\;\forall G\,,\;\;\left[ G\right] =\left[ A\right] -\left[ a%
\right] +\sum_{i}\left[ a_{i|J+1}\right] \,.  \label{3.3a}
\end{eqnarray}%
{\large That implies}%
\begin{equation}
F_{A}=\hat{U}_{A}^{B}\bar{F}_{B}\,,\;\bar{F}_{B}=\hat{V}_{B}^{A}F_{A}\,,\;\;%
\hat{U}_{A}^{B}\hat{V}_{B}^{C}=\delta _{A}^{C}\,,  \label{3.3b}
\end{equation}%
$\;$ {\large where} $\hat{U}$ {\large and} $\hat{V}$ {\large are LO}.

Let us make some comments to the Theorem 2. The canonical form (\ref{3.3})
of the singular ELE differs from that (\ref{2.28b}) of the nonsingular ELE.
As was demonstrated in the previous Sect., in the latter case the spectrum
of the orders of the variables $q^{a_{i}}$ in the canonical form extends
from $i+\mu _{\min }$ to $i+J\,.$ In the singular case, we have to admit
(and one can see this on specific examples) the spectrum extends from $0$ to 
$i+J\,.$ Under such a supposition we can justify by the induction the
structure (\ref{3.3}) of the canonical form. One can see from (\ref{3.4})
that each group of the indices $a_{i}$ is divided in subgroups $%
a_{i}\rightarrow a_{i|\sigma }\,,$ $\sigma =-I,...,J+1$. In the canonical
form the singular ELE are solved with respect to the highest-order
derivatives $q^{a_{i|\sigma }\left[ i+\sigma \right] }\,,\;\sigma =-I,...,J$%
\thinspace , ($\left[ a_{i|\sigma }\right] =0\;$for $i+\sigma <0)$. There
are no equations for the coordinates $q^{a_{i|J+1}}\,$. These coordinates
enter the set (\ref{3.3}) as arbitrary functions of time. They are gauge
coordinates according to the general definition. As in the nonsingular case,
it is supposed that no coordinate $q^{a_{k|\sigma }}$ in the function $%
\varphi ^{a_{i|\sigma }}$ has an order greater than $k+\sigma $ (the proper
order plus $\sigma $). Besides, the order of the coordinates $q^{b_{k|\sigma
_{-}}}$ in the function $\varphi ^{a_{i|\sigma }}$ has to be less than $%
k+\sigma _{-}$ .

We are going to prove the Theorem 2 by induction w.r.t. $\mathcal{N}=I+J$.
To this end, we consider first equations of lower orders, then we use an
induction to prove the general case.

\subsection{Equations of lower orders}

Remark that the case$\ \mathcal{N}=0\,$implies$\;I=J=0$ and the set (\ref%
{3.1}) is reduced to form%
\begin{equation}
F_{A}\left( q\right) =0,\;\;q=\left( q^{a}\right) \,.  \label{3.5b}
\end{equation}%
Here the Theorem 2 holds true by virtue of the Lemma 3 from the Appendix.

Let $\mathcal{N}=1.$ That implies either $I=1,$ \ $J=0$ or $I=0,\;$ $J=1$ .
Consider, for example, the first case. Here $\left( i=0,1,\;\mu =0\right) $
and the set (\ref{3.1}) reads 
\begin{equation}
F_{A}\left( q^{a_{0}},q^{a_{1}},\dot{q}^{a_{1}}\right) =0\,,\;\left[ a_{1}%
\right] >0,\;\,\left[ a_{0}\right] \geq 0\,.  \label{3.26}
\end{equation}%
In the case under consideration the supposition (\ref{3.2}) reads: 
\begin{equation}
\mathrm{rank}\,\frac{\partial F_{A}}{\partial q^{a_{i}\left[ i\right] }}=r\,.
\label{3.26a}
\end{equation}%
Then there exists a division of the indices: $\;A=\left(
A_{/1},A_{/2}\right) $,\ $a_{i}=\left( a_{i/1},a_{i/2}\right) $,\ $\left[
A_{/1}\right] =\left[ a_{0/1}\right] +\left[ a_{0/1}\right] =r$, \ such that

\begin{equation*}
\det \,\left| \frac{\partial F_{A_{/1}}}{\partial q^{a_{i/1}\left[ i\right] }%
}\right| \neq 0\,.
\end{equation*}%
Thus, we may solve the equations $F_{A_{/1}}=0$ with respect to $q^{a_{i/1}%
\left[ i\right] }$\thinspace ,%
\begin{equation}
F_{A_{/1}}=0\Longleftrightarrow q^{a_{i/1}\left[ i\right] }=\phi
^{a_{i/1}}\left( q^{b_{i/1}\left[ i-1\right] },q^{b_{i/2}\left[ i-1\right]
},q^{b_{i/2}\left[ i\right] }\right) \,.  \label{3.27}
\end{equation}%
Then we exclude the arguments $q^{a_{i/1}\left[ i\right] }$ from the
functions $F_{A_{/2}}$ by the help of (\ref{3.27}),%
\begin{equation*}
\bar{F}_{A_{/2}}=\left. F_{A_{/2}}\right| _{^{_{q^{a_{i/1}\left[ i\right]
}=\phi ^{a_{i/1}}}}}=\bar{F}_{A_{/2}}\left( q^{a_{1}}\right) \,.
\end{equation*}%
By virtue of the Lemma 2 from the Appendix, the functions $\bar{F}_{A_{/2}}$
depend on $q^{a_{1}}$ only . Thus, we have the equivalence\footnote{%
Here, and in what follows, we use Lemma 1 to justify the equivalence.}%
\begin{equation}
F_{A}\sim \bar{F}_{A}=\left( 
\begin{array}{l}
F_{A_{/1}}\left( q^{a_{0}},q^{a_{1}},\dot{q}^{a_{1}}\right) \\ 
\bar{F}_{A_{/2}}\left( q^{a_{1}}\right)%
\end{array}%
\right) .  \label{3.27a}
\end{equation}%
Now we suppose that the matrix $\partial \bar{F}_{A_{/2}}/\partial q^{a_{1}}$
has a constant rank. Therefore (see Lemma 3)%
\begin{equation*}
\bar{F}_{A_{/2}}\sim \left( 
\begin{array}{c}
q^{\underline{a}_{1}}-\varphi ^{\underline{a}_{1}}\left( q^{\bar{a}%
_{1}}\right) \\ 
0_{G_{1}}%
\end{array}%
\right) \,,\;\;a_{1}=\left( \underline{a}_{1},\bar{a}_{1}\right) \,.
\end{equation*}%
Let us exclude the arguments $q^{\underline{a}_{1}}$, $\dot{q}^{\underline{a}%
_{1}}$ from the functions $F_{A_{/1}}$, by the help of the equations $q^{%
\underline{a}_{1}}=\varphi ^{\underline{a}_{1}}\left( q^{\bar{a}_{1}}\right)
,$%
\begin{equation*}
\overset{1}{F}_{A_{/1}}\left( q^{a_{0}},q^{\bar{a}_{1}},\dot{q}^{\bar{a}%
_{1}}\right) =\left. F_{A_{/1}}\right| _{q^{\underline{a}_{1}}=\varphi ^{%
\underline{a}_{1}}}\,.
\end{equation*}%
Then the equivalence holds true: 
\begin{equation}
F\sim \overset{1}{F}=\left( 
\begin{array}{c}
\overset{1}{F}_{A_{/1}}\left( q^{a_{0}},q^{\bar{a}_{1}},\dot{q}^{\bar{a}%
_{1}}\right) \\ 
q^{\underline{a}_{1}}-\varphi ^{\underline{a}_{1}}\left( q^{\bar{a}%
_{1}}\right) \\ 
0_{G_{1}}%
\end{array}%
\right) ,\;\;a=\left( a_{0},a_{1}\right) \,,\;\;a_{1}=\left( \underline{a}%
_{1},\bar{a}_{1}\right) \,.  \label{3.28}
\end{equation}%
The set of functions $\overset{1}{F}$ has the same structure as the initial
set $F$. However, the number of the nonzero functions $\overset{1}{F}$ is
less than the number of the functions $F.$ Moreover, some of the functions $%
\overset{1}{F}$ depend linearly on a part of the variables. That is why the
supposition of the type (\ref{3.26a}) for the functions $F^{\left( 1\right)
} $ is reduced to the supposition about the rank of the matrix $\partial 
\overset{1}{F}_{A_{/1}}/\partial \left( q^{a_{0}},\dot{q}^{\bar{a}%
_{1}}\right) $ . Accepting the latter supposition we apply the above
reduction procedure to the functions $\overset{1}{F}$ and so on. After the $%
i $-th stage we have the following equivalence: 
\begin{equation*}
F\sim \overset{i}{F}=\left\{ 
\begin{array}{c}
\overset{i}{F}_{A_{/i}}\left( q^{a_{0}},q^{\bar{a}_{i}},\dot{q}^{\bar{a}%
_{i}},\right) \\ 
q^{\underline{a}_{i}}-\varphi ^{\underline{a}_{i}}\left( q^{\bar{a}%
_{i}}\right) \\ 
0_{G_{i}}%
\end{array}%
\right. ,\;a=\left( a_{0},a_{1}\right) \,,\;\;a_{1}=\left( \underline{a}_{i},%
\bar{a}_{i}\right) \,.
\end{equation*}%
The procedure ends at a $k$-th stage when%
\begin{equation*}
\mathrm{rank\,}\frac{\partial \overset{k}{F}_{A_{/k}}}{\partial \left( \dot{q%
}^{\bar{a}_{k}},q^{a_{0}}\right) }=\left[ A_{/k}\right] \,.
\end{equation*}%
Then there exists a division of the indices $\bar{a}_{k}=\left(
a_{1|0},a_{g_{1}}\right) $,\ $a_{0}=\left( a_{0|0},a_{g_{0}}\right) $,\ $%
\left[ a_{1|0}\right] +\left[ a_{0|0}\right] =\left[ A_{/k}\right] $, such
that%
\begin{equation*}
\det \,\frac{\partial \overset{k}{F}_{A_{/k}}}{\partial (\dot{q}%
^{a_{1|0}},q^{a_{0|0}})}\neq 0\,\Longrightarrow \overset{k}{F}_{A_{/k}}\sim
\left( 
\begin{array}{c}
\dot{q}^{a_{1|0}}-\varphi ^{a_{1|0}}\left(
q^{a_{1|0}},q^{a_{g_{0}}},q^{a_{g_{1}}},\dot{q}^{a_{g_{1}}}\right) \\ 
q^{a_{0|0}}-\varphi ^{a_{0|0}}\left( q^{a_{1|0}},q^{a_{g_{0}}},q^{a_{g_{1}}},%
\dot{q}^{a_{g_{1}}}\right)%
\end{array}%
\right) \,.
\end{equation*}%
Denoting $\underline{a}_{k}\equiv a_{1|-1}\,$,\ $G=G_{k}$, such that $%
a=\left( a_{1|-1},a_{0|0},a_{1|0},a_{g}\right) $,\ and $a_{g}=\left(
a_{g_{0}},a_{g_{1}}\right) $,\ $\left[ G\right] =\left[ A\right] -\left[ a%
\right] +\left[ a_{g}\right] $, we get finally the equivalence: 
\begin{equation}
F\sim \left( 
\begin{array}{c}
\dot{q}^{a_{1|0}}-\varphi ^{a_{1|0}}\left(
q^{a_{1|0}},q^{a_{g_{0}}},q^{a_{g_{1}}},\dot{q}^{a_{g_{1}}}\right) \\ 
q^{a_{0|0}}-\varphi ^{a_{0|0}}\left( q^{a_{1|0}},q^{a_{g_{0}}},q^{a_{g_{1}}},%
\dot{q}^{a_{g_{1}}}\right) \\ 
q^{a_{1|-1}}-\varphi ^{a_{1|-1}}\left( q^{a_{1|0}},q^{a_{g_{1}}}\right) \\ 
0_{G}%
\end{array}%
\right) \,.  \label{3.29}
\end{equation}%
Here $q^{a_{g}}=\left( q^{a_{g_{0}}},q^{a_{g_{1}}}\right) $ are gauge
coordinates. Thus, the Theorem 2 holds true in this case.

The case $I=0,\;$ $J=1$ $\left( i=0,\;\mu =0,1\right) $ corresponds to the
equations of the form 
\begin{equation}
F_{A_{1}}\left( q^{a_{1}},\dot{q}^{a_{1}}\right) =0\,,\;F_{A_{0}}\left(
q^{a_{1}}\right) =0\,.  \label{3.15}
\end{equation}%
Such equations present a particular case ($\left[ a_{0}\right] =0$) of the
equations $\bar{F}_{A}=0$ with the LF $\bar{F}_{A}$ defined in (\ref{3.27a}%
). The reduction procedure for the latter case was considered above. It
leads to the following equivalence%
\begin{equation*}
F\sim \left( 
\begin{array}{c}
\dot{q}^{a_{|1}}-\varphi ^{a_{|1}}\left( q^{a_{|1}},q^{a_{g}},\dot{q}%
^{a_{g}}\right) \\ 
q^{a_{|0}}-\varphi ^{a_{|0}}(q^{a_{|1}},q^{a_{g}}) \\ 
0_{G}%
\end{array}%
\right) \,,\;\;a=\left( a_{|0},a_{|1},a_{g}\right) \,,\;\left[ G\right] =%
\left[ A\right] -\left[ a\right] +\left[ a_{g}\right] .
\end{equation*}%
Here $q^{a_{g}}$ are the gauge coordinates. Thus, the Theorem holds true in
this case as well.

\subsection{Equations of arbitrary orders}

We have verified that the Theorem 2 holds true for $\mathcal{N}=0,1$. Now we
are going to prove the theorem for $\mathcal{N}=I+J=K$ (where $K$ is some
fixed number)\ supposing that the theorem holds true for any $\mathcal{N}<K$
.

At the first step we consider the set 
\begin{equation}
F_{A_{\mu }}^{\left[ J-\mu \right] }\left( \cdots q^{a_{i}\left[ i+J\right]
}\right) =0\,,\;\;i=0,1,...,I\,,\;\mu =0,...,J\,,  \label{3.6}
\end{equation}%
which is obtained from the initial set (\ref{3.1}) by substituting the
constraints by the corresponding consistency conditions (conditions obtained
from the constraints $F_{A_{\mu }}$ by $J-\mu $ time differentiations).
According to the supposition (\ref{3.2}), there exists a division of the
indices $A_{\mu }$ and $a_{i}\;$as:$\;A_{\mu }=\left( A_{\mu /1},A_{\mu
/2}\right) ,$ $a_{i}=\left( a_{i/1},a_{i/2}\right) ,\;$ $\sum_{\mu }\left[
A_{\mu /1}\right] =\sum_{i}\left[ a_{i/1}\right] =r$, such that:

\begin{equation}
\det \,\left| \frac{\partial F_{A_{\mu /1}}^{\left[ J-\mu \right] }}{%
\partial q^{a_{i/1}\left[ i+J\right] }}\right| \neq 0\,.  \label{3.5}
\end{equation}%
Thus, we may solve the equations $F_{A_{\mu /1}}^{\left[ J-\mu \right] }=0$
with respect to the derivatives $q^{a_{i/1}\left[ i+J\right] }\,.$ Namely, 
\begin{equation}
\;F_{A_{\mu /1}}^{\left[ J-\mu \right] }=0\Longleftrightarrow q^{a_{i/1}%
\left[ i+J\right] }=\varphi ^{a_{i/1}}\left( \cdots q^{b_{j/1}\left[ j+J-1%
\right] },\cdots q^{b_{j/2}\left[ j+J\right] }\right) \,.  \label{3.7a}
\end{equation}%
Now we pass from the functions $F_{A_{J/2}}$ to the ones $\bar{F}_{A_{J/2}}$
excluding the arguments $q^{b_{i/1}\left[ i+J\right] }$ from the former, 
\begin{equation}
\bar{F}_{A_{J/2}}=\left. F_{A_{J/2}}\right| _{f=0}\,=\bar{F}_{A_{J/2}}\left(
\cdots q^{b_{i}\left[ i+J-1\right] }\right) \,.  \label{3.8c}
\end{equation}%
The fact that the functions $\bar{F}_{A_{J/2}}$ do not depend on both $%
q^{b_{i/1}\left[ i+J\right] }$ and $q^{b_{i/2}\left[ i+J\right] }\,$\ is
based on the Lemma 2 from the Appendix. Thus, we have the equivalence (see
Lemma 1 from the Appendix) 
\begin{equation}
F_{A}\sim \left( 
\begin{array}{c}
F_{A_{J/1}} \\ 
\bar{F}_{A_{J/2}} \\ 
F_{A_{\nu }}\,,\;\nu =0,...,J-1%
\end{array}%
\right) \sim \left( 
\begin{array}{c}
F_{A_{J/1}} \\ 
F_{A^{\prime }}^{\prime }\,,\;%
\end{array}%
\right) \,,  \label{3.9}
\end{equation}%
where 
\begin{equation}
F_{A^{\prime }}^{\prime }=\left( F_{A_{\nu }^{\prime }}^{\prime }\,,\;\nu
=0,...,J-1\right) =\left\{ 
\begin{array}{c}
F_{A_{\nu }}\left( \cdots q^{b_{i}\left[ i+\nu \right] }\right) \,,\;\nu
=0,...,J-2 \\ 
F_{A_{J-1}^{\prime }}^{\prime }\left( \cdots q^{b_{i}\left[ i+J-1\right]
}\right) =\left\{ 
\begin{array}{c}
F_{A_{J-1}} \\ 
\bar{F}_{A_{J/2}}%
\end{array}%
\right.%
\end{array}%
\right. \,\,.  \label{3.10a}
\end{equation}

Let us turn to the functions $F_{A_{\nu }^{\prime }}^{\prime }$ . They have
the same structure as in (\ref{3.1}) and correspond to the case $\mathcal{N}%
=I+J<K.$ In accordance with the induction hypothesis, supposing, in
particular, that the matrix 
\begin{equation*}
M_{A_{\nu }^{\prime }\,a_{i}}^{\prime }=\frac{\partial F_{A_{\nu }^{\prime
}}^{\prime }}{\partial q^{a_{i}\left[ i+\nu \right] }}
\end{equation*}%
has a constant rank in the consideration point the following equivalence
holds true:%
\begin{eqnarray}
&&F_{A^{\prime }}^{\prime }\sim \left( 
\begin{array}{c}
q^{a_{i|\sigma }\left[ i+\sigma \right] }-\varphi ^{a_{i|\sigma }}\left(
\cdots q^{b_{j|\sigma _{-}}\left[ j+\sigma _{-}-1\right] },\cdots
q^{b_{j|\sigma _{+}}\left[ j+\sigma \right] },\cdots q^{b_{j|J}\left[
j+\sigma \right] }\right) \\ 
0_{G^{\prime }}%
\end{array}%
\right) \,,  \notag \\
&&i,j=0,1,...,I\;,\;\;\left[ G^{\prime }\right] =\left[ A^{\prime }\right] -%
\left[ a\right] +\sum_{i}\left[ a_{i|J}\right] \;,  \notag \\
&&\sigma =-I,...,J-1,\;\;-I\leq \sigma _{-}\leq \sigma \,,\;\;\sigma +1\leq
\sigma _{+}\leq J-1\,.  \label{3.10b}
\end{eqnarray}%
Taking into account (\ref{3.9}), we obtain 
\begin{eqnarray}
&&F\sim \left( 
\begin{array}{c}
F_{A_{J/1}}\left( \cdots q^{b_{i}\left[ i+J\right] }\right) \\ 
q^{a_{i|\sigma }\left[ i+\sigma \right] }-\varphi ^{a_{i|\sigma }}\left(
\cdots q^{b_{j|\sigma _{-}}\left[ j+\sigma _{-}-1\right] },\cdots
q^{b_{j|\sigma _{+}}\left[ j+\sigma \right] },\cdots q^{b_{j|J}\left[
j+\sigma \right] }\right) \\ 
0_{G^{\prime }}%
\end{array}%
\right) \,,  \notag \\
&&i,j=0,1,...,I\;,\;\sigma =-I,...,J-1,\;-I\leq \sigma _{-}\leq \sigma
\,,\;\sigma +1\leq \sigma _{+}\leq J-1\,.  \label{3.11a}
\end{eqnarray}%
Now we pass from the functions $F_{A_{J/1}}$ to the ones $\bar{F}_{A_{J/1}}$
excluding the arguments $q^{a_{i|\sigma }\left[ p_{i}\right] },\;$ $%
p_{i}\geq i+\sigma ,$ $\sigma =-I,...,J-1$ from the former. As a result, the
following equivalence takes place: 
\begin{equation}
F\sim \tilde{F}=\left( 
\begin{array}{c}
\bar{F}_{A_{J/1}}\left( \cdots q^{b_{i|J}\left[ i+J\right] },\cdots
q^{b_{i|\sigma }\left[ i+\sigma -1\right] }\right) \\ 
q^{a_{i|\sigma }\left[ i+\sigma \right] }-\varphi ^{a_{i|\sigma }}\left(
\cdots q^{b_{j|\sigma _{-}}\left[ j+\sigma _{-}-1\right] },\cdots
q^{b_{j|\sigma _{+}}\left[ j+\sigma \right] },\cdots q^{b_{j|J}\left[
j+\sigma \right] }\right) \\ 
0_{G^{\prime }}%
\end{array}%
\right) \,.  \label{3.11b}
\end{equation}

The functions $\tilde{F}$ have the same structure as in (\ref{3.1}),
however, they depend linearly on a part of highest-order derivatives. Here
the supposition of the rank for the matrix 
\begin{equation}
\frac{\partial \tilde{F}_{A}}{\partial (q^{a_{i|J}\left[ i+J\right]
},q^{a_{i|\sigma }\left[ i+\sigma \right] })}\,,\;A=\left(
A_{J/1}\,,a_{i|\sigma }\,,G^{\prime }\right)  \label{3.12}
\end{equation}%
is equivalent to the same supposition for the matrix 
\begin{equation}
\frac{\partial \bar{F}_{A_{J/1}}}{\partial q^{b_{i|J}\left[ i+J\right] }}\,.
\label{3.13}
\end{equation}%
Let this rank be equal to $\left[ {A_{J/1}}\right] $. In this case there
exists a final division of indices, 
\begin{equation*}
a_{i|J}\rightarrow \left( a_{i|J},a_{i|J+1}\right) \,\;\mathrm{with}%
\;\;[a_{i|J}]=[A_{J/1}]\,,
\end{equation*}%
such that the equations $\bar{F}_{A_{J/1}}=0$ can be solved with respect to
the derivatives $q^{a_{i|J}\left[ i+J\right] }$ and we obtain, instead of
the two first lines of (\ref{3.11b}), the following expressions: 
\begin{eqnarray}
&&q^{a_{i|J}\left[ i+J\right] }-\varphi ^{a_{i|J}}\left( \cdots q^{b_{j|J}%
\left[ j+J-1\right] },\cdots q^{b_{j|\sigma }\left[ j+\sigma -1\right]
},\cdots q^{b_{j|J+1}\left[ j+J\right] }\right) ,  \notag \\
&&q^{a_{i|\sigma }\left[ i+\sigma \right] }-\varphi ^{a_{i|\sigma }}\left(
\cdots q^{b_{j|\sigma _{-}}\left[ j+\sigma _{-}-1\right] },\cdots
q^{b_{j|\sigma _{+}}\left[ j+\sigma \right] },\cdots q^{b_{j|J}\left[
j+\sigma \right] }\right) ,  \notag \\
&&i,j=0,1,...,I\;,\;\sigma =-I,...,J-1,\;-I\leq \sigma _{-}\leq \sigma
\,,\;\sigma +1\leq \sigma _{+}\leq J-1\,.  \notag
\end{eqnarray}%
Now, let us put together the first two entries of $\varphi ^{a_{i|J}}$ as $%
\cdots q^{b_{j|\sigma }\left[ j+\sigma -1\right] },\;-I\leq \sigma \leq J$
and remind that for $\sigma =J$ no corresponding $\sigma _{+}$ occurs.
Furthermore, let us replace the last entry of $\varphi ^{a_{i|\sigma }}$ as
follows: $\cdots q^{b_{j|J}\left[ j+\sigma \right] }\rightarrow \cdots
q^{b_{j|J}\left[ j+\sigma \right] }\cdots q^{b_{j|J+1}\left[ j+\sigma \right]
},\;\;-I\leq \sigma \leq J-1$, then we get the missing contribution to $%
\sigma _{+}$ for the case under consideration. So, we end up exactly with
Eq. (\ref{3.3}) and the Theorem 2 is proved.

If the rank is less than $\left[ A_{J/1}\right] $ then the above procedure
is applied to the functions $\bar{F}_{A_{J/1}}$. Doing that we lower the
number of the equations that are not yet reduced to the canonical form (the
equations of the type $\bar{F}_{A_{J/1}}=0$ ). Remark that such a diminution
does not happen at the first stage if $\left[ A_{J/2}\right] =0$ . At a
certain stage the procedure does not lower the number of the above mentioned
equations. This can happen when the rank of the matrix of the type (\ref%
{3.13}) is maximal, i.e. is equal to the number of the functions of the type 
$\bar{F}_{A_{J/1}}\,$. In such a case we may reduce them to the canonical
form as was said above. This can also happen when we do not obtain the
functions of the type $\bar{F}_{A_{J/1}}$ in the reduction procedure. That
means that already at the previous step the set is reduced to the case $%
\mathcal{N}=K-1,$ i.e. the possibility of the reduction to the canonical
form is proved.

Finally we stress that the reduction procedure is formulated for sets of
equations of the type (\ref{3.1}) (the ELE are a particular case of such
sets). The procedure holds true under certain suppositions of ranks. These
suppositions demand various Jacobi matrices of the type $\partial
F_{s}/\partial q^{a\left[ l\right] }$ to have constant ranks in the vicinity
of the consideration point. Here $F_{s}=0$ are equations obtained to a given
stage of the procedure and $q^{a\left[ l\right] }$ are highest-order
derivatives in these equations. It is important to realize that proving the
equivalence (\ref{3.3a}) we prove at the same time the locality of the
operators $\hat{U}$ and $\hat{V}$ from (\ref{3.3b}). In fact, the latter
proof is provided by the applicability of the Lemmas from the Appendix.

\section{Gauge identities and action symmetries}

It was demonstrated above that in the general case of singular ELE the
number of the equations in the canonical form is less than the number of the
equations in the initial set of the differential equations. This reduction
is related to the fact that in the canonical form we retain the independent
equations only, whereas the initial equations may be dependent. The
dependence of the equations in the initial set may be treated as the
existence of some identities between the initial equations. The identities
between the ELE imply the existence of gauge transformations of the
corresponding action. Below we discuss this interrelationship in detail.

First, we introduce some relevant definitions: The relation of the form 
\begin{equation}
\hat{R}^{a}F_{a}\equiv 0\,,  \label{4.1}
\end{equation}%
where $\hat{R}^{a}$ are some LO, and $F_{a}\left( q^{\left[ l\right]
}\right) $ are some LF, is called the\emph{\ }identity between the equations 
$F_{a}\left( q^{\left[ l\right] }\right) =0.$ The identity sign $\equiv $
means that the left hand side of (\ref{4.1}) is zero for any arguments $q^{%
\left[ l\right] }\,.$

Any set $\mathbf{\hat{R}}=\left( \hat{R}^{a}\right) $ of LO that obeys the
relation (\ref{4.1}) is called the\emph{\ }generator of an identity.\emph{\ }%
Whenever $\mathbf{\hat{R}}$ is a generator than $\hat{n}\mathbf{\hat{R}}$
with some LO $\hat{n}$ is a generator as well. Any linear combination $\hat{n%
}^{i}\mathbf{\hat{R}}_{i}$ of some generators $\mathbf{\hat{R}}_{i}\,\ $with
operator coefficients $\hat{n}^{i}$ is a generator.

A generator $\mathbf{\hat{R}}$ will be called nontrivial if the relation%
\footnote{%
By $\hat{O}(F)$ we denote LO of the form (\ref{2.10}) with all the LF $%
u_{Aa}^{k}=O(F)\,,$ where%
\begin{equation*}
\;\;\;\left. O(F)\right| _{F=0}=0\,.
\end{equation*}%
} $\hat{n}\mathbf{\hat{R}}=\hat{O}\left( F\right) $ can only be provided by
a LO $\hat{n}$ of the form $\hat{n}=\hat{O}\left( F\right) \,$.

A set of generators $\mathbf{\hat{R}}_{i}$ will be called independent if the
relation $\hat{n}^{i}\mathbf{\hat{R}}_{i}=\hat{O}\left( F\right) $ can only
be provided by $\hat{n}^{i}$ of the form $\hat{n}^{i}=\hat{O}\left( F\right)
\,.$ Identities generated\ by independent generators will be called
independent.

Note that for any set of LF $F_{a}$ , there always exist trivial generators.
Namely, the generators $\mathbf{\hat{R}}_{\mathrm{triv}}=\left( \hat{R}_{%
\mathrm{triv}}^{a}\right) =\hat{O}\left( F\right) $ of the form 
\begin{equation}
\hat{R}_{\mathrm{triv}}^{a}=\sum_{k,l}F_{b}^{\left[ k\right] }u^{bk|al}\frac{%
d^{l}}{dt^{l}}\,,\;u^{bk|al}=-u^{al|bk}\,,  \label{4.4}
\end{equation}%
with arbitrary antisymmetric LF $u^{bk|al}\;$obviously lead to the
identities (\ref{4.1}). These identities are not, however, connected to the
mutual dependence of the functions $F_{a}$ $.$

An independent set of generators $\mathbf{\hat{R}}_{g}$ is complete whenever
any generator $\mathbf{\hat{R}}$ can be represented in the form $\mathbf{%
\hat{R}}=\hat{\lambda}^{g}\mathbf{\hat{R}}_{g}+\mathbf{\hat{R}}_{\mathrm{triv%
}}$ with some LO $\hat{\lambda}^{g}.$ Any two complete sets of independent
generators $\mathbf{\hat{R}}_{g}$ and $\mathbf{\hat{R}}_{g}^{\prime }$ are
related as $\mathbf{\hat{R}}_{g}^{\prime }=\hat{U}_{g}^{g^{\prime }}\mathbf{%
\hat{R}}_{g^{\prime }}+\mathbf{\hat{R}}_{\mathrm{triv}}\,,$where $\hat{U}$
is an invertible LO.

Supposing now that $F_{a}$ in Eq. (\ref{4.1}) are functional derivatives of
an action, $F_{a}=\delta S/\delta q^{a},$ such that $F_{a}=0$ are ELE. Let
the functions $F_{a}$ obey all the necessary suppositions of ranks such that
ELE can be reduced to the canonical form (\ref{3.3}). Let us write here this
canonical form as follows\footnote{%
Here, we do not distinguish possible different proper orders of the
coordinates.},%
\begin{equation}
f^{\alpha }=q^{\alpha \left[ l_{\alpha }\right] }-\varphi ^{\alpha }\left(
\cdots q^{\alpha \left[ l_{\alpha }-1\right] };\cdots q^{g\left[ l_{g}\right]
}\right) =0\,,\;a=\left( \alpha ,g\right) \,,  \label{4.5a}
\end{equation}%
where $q^{g}$ are gauge coordinates. Moreover, according to the Theorem 2,
there exists the equivalence%
\begin{equation}
F_{a}\sim \bar{F}_{a}=\left( 
\begin{array}{l}
f^{\alpha } \\ 
0_{g}%
\end{array}%
\right) \Longrightarrow F_{a}=\hat{U}_{a}^{b}\bar{F}_{b}\,,\;\bar{F}_{a}=%
\hat{V}_{a}^{b}F_{b}\,,\;\hat{U}_{a}^{b}\hat{V}_{b}^{c}=\delta _{a}^{c}\,,
\label{4.5}
\end{equation}%
where $\hat{U}$ and $\hat{V}$ are LO. Now we may consider the identity (\ref%
{4.1}) as an equation for finding the general form for the generator $%
\mathbf{\hat{R}}$ . Using (\ref{4.5}) we transform this problem to the one
for finding the operators $\hat{\xi}^{a},$%
\begin{equation}
\hat{\xi}^{a}\bar{F}_{a}\equiv 0\,,\;\;\hat{R}^{a}=\hat{\xi}^{b}\hat{V}%
_{b}^{a}\,.  \label{4.6}
\end{equation}%
Using the explicit form (\ref{4.5}) of the functions $\bar{F}_{a}$ , we get $%
\hat{\xi}^{a}=\left( 
\begin{array}{cc}
\hat{\xi}^{\alpha } & \hat{\xi}^{g}%
\end{array}%
\right) ,\;a=\left( \alpha ,g\right) $ , where $\hat{\xi}^{\alpha }$ obey
the equation 
\begin{equation}
\hat{\xi}^{\alpha }f^{\alpha }\equiv 0\,,  \label{4.7}
\end{equation}%
and $\hat{\xi}^{g}$ is a set of arbitrary LO. Since the functions $f$ have
the canonical form (\ref{4.5}), any solution of the equation (\ref{4.7}) is
presented by trivial generators of the form 
\begin{equation}
\hat{\xi}^{\alpha }=\hat{\xi}_{\mathrm{triv}}^{\alpha }=\sum_{k,l}\left( 
\frac{d^{l}}{dt^{l}}f_{\alpha ^{\prime }}\right) u^{l\alpha ^{\prime
}|k\alpha }\frac{d^{k}}{dt^{k}}\,,\;\;u^{l\alpha ^{\prime }|k\alpha
}=-u^{k\alpha |l\alpha ^{\prime }}\,,  \label{4.8}
\end{equation}%
where $u^{l\alpha ^{\prime }|k\alpha }$ are arbitrary antisymmetric LF. To
demonstrate that we present the generators $\hat{\xi}^{\alpha }$ as $\hat{\xi%
}^{\alpha }=\sum_{k=0}^{K}\xi ^{\alpha k}d^{k}/dt^{k}\,,$ where $\xi
^{\alpha k}$ are some LF. Then, in the equation (\ref{4.7}), we pass from
the variables $q^{\alpha \left[ k\right] },\;q^{g\left[ l\right] },$ $%
\;k,l=0,1,...$ to ones $q^{\alpha \left[ k_{\alpha }\right] },\;f_{\alpha }^{%
\left[ l\right] },\;q^{g\left[ l\right] },$ $k_{\alpha }=0,1,...,l_{\alpha
}-1,\;$ $l=0,1,...$ . Such a variable change is not singular. In terms of
the new variables, the equation (\ref{4.7}) reads%
\begin{equation*}
\sum_{k=0}^{K}\xi ^{\alpha k}f_{\alpha }^{\left[ k\right] }=0\;,\;K<\infty
\;.
\end{equation*}%
Its general solution is well known 
\begin{equation*}
\xi ^{\alpha k}=\sum_{l}f_{\alpha ^{\prime }}^{\left[ l\right] }u^{l\alpha
^{\prime }|k\alpha }\,,\;u^{l\alpha ^{\prime }|k\alpha }=-u^{k\alpha
|l\alpha ^{\prime }}\;.
\end{equation*}%
Now we can write the general solution of the equation (\ref{4.6}) as: 
\begin{equation}
\hat{\xi}^{a}=\hat{\xi}^{g}\delta _{g}^{a}+\hat{\xi}_{\mathrm{triv}}^{a},\;\;%
\hat{\xi}_{\mathrm{triv}}^{a}=\sum_{k,l}\left( \frac{d^{l}}{dt^{l}}\bar{F}%
_{b}\right) u^{lb|ka}\frac{d^{k}}{dt^{k}},\;\;u^{lb|ka}=-u^{ka|lb}.
\label{4.12}
\end{equation}%
Let $b=\left( \alpha ^{\prime },g^{\prime }\right) ,\;$ $a=\left( \alpha
,g\right) $ in (\ref{4.12}) . Then $u^{lg^{\prime }|k\alpha }$, $u^{l\alpha
^{\prime }|kg}=-u^{kg|l\alpha ^{\prime }}$ and $u^{lg^{\prime }|kg}$ are
arbitrary LF\ (e.g., they can be selected to be zero). Indeed, the functions 
$u^{lg^{\prime }|k\alpha }$ and $u^{lg^{\prime }|kg}$ do not enter the
expressions for the generators $\hat{\xi}^{a}$. Besides, terms with $%
u^{l\alpha ^{\prime }|kg}$ affect only the generators $\hat{\xi}^{g}$, which
are arbitrary by the construction. Respectively, the general solution of the
equation (\ref{4.1}) reads: 
\begin{equation}
\mathbf{\hat{R}}=\hat{\xi}^{g}\mathbf{\hat{R}}_{g}+\mathbf{\hat{R}}_{\mathrm{%
triv}}\;,\;\;\mathbf{\hat{R}}_{g}=(\hat{R}_{g}^{a}=\delta _{g}^{b}\hat{V}%
_{b}^{a}=\hat{V}_{g}^{a})\,,  \label{4.13}
\end{equation}%
and 
\begin{equation*}
\hat{R}_{\mathrm{triv}}^{a}=\hat{\xi}_{\mathrm{triv}}^{b}\hat{V}%
_{b}^{a}=\sum_{k,l}\left[ \frac{d^{l}}{dt^{l}}\left( \hat{V}%
_{b}^{c}F_{c}\right) \right] u^{lb|kd}\frac{d^{k}}{dt^{k}}\hat{V}%
_{d}^{a}=\sum_{k,l}\left( \frac{d^{l}}{dt^{l}}F_{b}\right) T^{lb|ka}\frac{%
d^{k}}{dt^{k}}\,,
\end{equation*}%
where $T^{lb|ka}=-T^{ka|lb}\,$ are some LF. The set of the generators $%
\mathbf{\hat{R}}_{g}=(\hat{R}_{g}^{a}=\hat{V}_{g}^{a})$ is complete and it
is presented by LO. Moreover, these generators are independent. Indeed,
multiplying the equation $\hat{n}^{g}\hat{R}_{g}^{a}=\hat{O}\left( F\right) $
from the right by $\hat{U}_{a}^{b},$ we get: $\hat{n}^{g}\delta _{g}^{b}=%
\hat{O}\left( F\right) \Longrightarrow \hat{n}^{g}=\hat{O}\left( F\right) $.

Thus, there exist the following nontrivial identities between the ELE:%
\begin{equation}
\hat{R}_{g}^{a}\frac{\delta S}{\delta q^{a}}\equiv 0\,,\;g=1,...,r,
\label{4.14}
\end{equation}%
with generators $\mathbf{\hat{R}}_{g}$ that are LO. These identities are
called the\emph{\ }gauge identities. As is well known (see for example, \cite%
{GitTy90,HenTe92}), the existence of the gauge identities (\ref{4.14})
implies the existence of infinitesimal gauge transformations of the form 
\begin{equation}
q^{a}\rightarrow q^{a}+\delta q^{a}\,,\;\delta q^{a}=\left( \hat{R}%
^{T}\right) _{g}^{a}\epsilon ^{g}\;,  \label{4.15}
\end{equation}%
where $r$ parameters $\epsilon ^{g}=\epsilon ^{g}(t)$ are arbitrary
functions of time $t$. Remark that $\mathbf{\hat{R}}^{T}$ are LO as well.

Thus, it was demonstrated that for theories that obey appropriate
suppositions of the ranks there exists a constructive procedure of revealing
the gauge generators. For such theories all the generators are LO. The
number of the independent generators and, therefore, the number of the
independent gauge transformations is equal to the number of the gauge
coordinates in the ELE.

As a simple mechanical example, consider the action of the form\footnote{%
At $a\neq 0$ we have a finite-dimensional analog of the Proca action, and at 
$a=0$ we have the analog of the Maxwell action.}

\begin{equation}
S=\int Ldt\,,\;\;L=\frac{1}{2}\left( \dot{x}-y\right) ^{2}+\frac{a}{2}\left(
y^{2}-x^{2}\right) \,.  \label{e.7}
\end{equation}%
The corresponding ELE are: 
\begin{equation}
F_{1}=\ddot{x}-\dot{y}+ax=0,\;\;F_{2}=\dot{x}-\left( 1+a\right) y=0\,,
\label{e.8}
\end{equation}%
where $F_{2}=0$ is a constraint. The generalized Hessian reads:%
\begin{equation}
M=\left| 
\begin{array}{cc}
\frac{\partial ^{2}L}{\partial \dot{x}^{2}}=1 & \frac{\partial ^{2}L}{%
\partial \dot{x}\partial y}=-1 \\ 
\frac{\partial ^{2}L}{\partial y\partial \dot{x}}=-1 & \frac{\partial ^{2}L}{%
\partial y^{2}}=a+1%
\end{array}%
\right| =a\,.  \label{e.10}
\end{equation}

Let $a\neq 0\,,\;M\neq 0.$ In such a nonsingular case the reduction
procedure looks as follows: By the help of the consistency condition $\dot{F}%
_{2}=0\Longrightarrow \ddot{x}=\left( 1+a\right) \dot{y}\,,$ we eliminate $%
\ddot{x}$ from the first ELE. Thus, we get an equivalent set, which has the
canonical form,%
\begin{equation}
\dot{y}=-x\,,\;\;\,\dot{x}=\left( 1+a\right) y\;.  \label{e.11}
\end{equation}%
Another canonical form%
\begin{equation}
\ddot{x}=-\left( 1+a\right) x\,,\;\;y=\left( 1+a\right) ^{-1}\,\dot{x}\,,
\label{e.12}
\end{equation}%
we obtain eliminating $\dot{y}$ from the equation $F_{1}=0$ by the help of
the consistency condition $\dot{F}_{2}=0\Longrightarrow \dot{y}=\ddot{x}%
/\left( 1+a\right) $ $.$

Let $a=0\,.\;$The case is singular, $M=0\,,$ and rank of the Hessian matrix
is equal to$\;1$ . One can easily see that the equivalence%
\begin{equation*}
\left( 
\begin{array}{c}
F_{1} \\ 
F_{2}%
\end{array}%
\right) =\hat{U}\left( 
\begin{array}{c}
\dot{x}-y \\ 
0%
\end{array}%
\right) ,\;\hat{U}=\left( 
\begin{array}{cc}
d/dt & 1 \\ 
1 & 0%
\end{array}%
\right) ,\;\hat{U}^{-1}=\left( 
\begin{array}{cc}
0 & 1 \\ 
1 & -d/dt%
\end{array}%
\right)
\end{equation*}%
holds true. Then the canonical form of the ELE reads $\dot{x}=y$ and there
is a gauge identity%
\begin{equation*}
\hat{R}^{a}F_{a}\,\equiv 0\,,\;\;\hat{R}^{1}=1\,,\;\hat{R}^{2}=-d/dt\;.
\end{equation*}%
The operators transposed to $\hat{R}^{a}$ are $\left( \hat{R}^{T}\right)
^{a}=\left( \left( \hat{R}^{T}\right) ^{1}=1,\left( \hat{R}^{T}\right) ^{2}=%
\frac{d}{dt}\right) .$ Thus, at $a=0,$ the action (\ref{e.7}) is invariant
under the gauge transformation $x\rightarrow x+\epsilon \,,\;y\rightarrow y+%
\dot{\epsilon}\,.$ In the case under consideration, the ELE have two
canonical forms: $\dot{x}=y\;$and $y=\dot{x}$ .

\section{Concluding remarks}

We have formulated the reduction procedure which allows one to transform the
ELE to the canonical form as well as to establish possible gauge identities
between the equations. The latter part of the procedure can be considered as
a constructive way of finding all the gauge generators within the Lagrangian
formulation. At the same time, it is proven that, for local theories, all
the gauge generators are local in time operators. The canonical form of the
ELE reveals their hidden structure, in particular, it presents the spectrum
of possible initial data, and it allows one to separate coordinates into
nongauge and gauge ones. One ought also to remark that the reduction
procedure can be, in particular, treated as a procedure of finding
constraints in the Lagrangian formulation.

In that respect one can compare the reduction procedure with the well-known
Dirac procedure in the Hamiltonian formulation of constrained systems \cite%
{Dirac64,GitTy90,HenTe92}. Recall that the Dirac procedure is applicable to
the Hamilton equations with primary constraints, namely to equations of the
form%
\begin{equation}
F\left( \eta ,\dot{\eta}\right) =\dot{\eta}-\left\{ \eta \,,H^{\left(
1\right) }\right\} =0\,,\,\,\Phi ^{\left( 1\right) }\left( \eta \right)
=0\,,\;H^{\left( 1\right) }=H\left( \eta \right) +\lambda \Phi ^{\left(
1\right) }\left( \eta \right) \,.  \label{r.1}
\end{equation}%
Here $\eta =\left( q^{a},p_{a}\right) $ are phase-space variables; $\Phi
^{\left( 1\right) }\left( \eta \right) =0\;$are primary constraints, $%
\lambda $'s are Lagrange multipliers to the primary constraints, and $%
H^{\left( 1\right) }$ is the total Hamiltonian . Via $\{\cdot ,\cdot \}$ the
Poisson bracket is denoted. The aim of the procedure is to eliminate as many
as possible $\lambda $'s from the equations, to find all the constraints in
the theory. The procedure is based on the consistency conditions $\dot{\Phi}%
^{\left( 1\right) }=0$. Using the equations $F\left( \eta ,\dot{\eta}\right)
=0$, we may transform any consistency condition to the following form:%
\begin{equation*}
\dot{\Phi}^{\left( 1\right) }=\left\{ \Phi ^{\left( 1\right) },H^{\left(
1\right) }\right\} =0\,.
\end{equation*}%
From these equations one can define some $\lambda $'s as functions of $\eta $
and reveal some new constraints. Then the procedure has to be applied to the
latter constraints and so on.

The equations (\ref{r.1}) present a particular case of differential
equations considered in the present article (indeed, these equations are ELE
for a Hamiltonian action). Thus, our reduction procedure may be applied to
these equations. Namely, first one has to consider the equations $F_{A}=0,\;%
\dot{\Phi}^{\left( 1\right) }=0$ and select independent w.r.t. $\dot{\eta}$
equations. Since equations of the primary constraints are independent by
construction, we pass to the next step and solve the constraint equations $%
\Phi ^{\left( 1\right) }=0$ with respect to a part of the variables $\eta ,$
as $\Phi ^{\left( 1\right) }=0\;\rightarrow \;\eta _{1}-\varphi _{1}\left(
\eta _{2}\right) =0$. Then we exclude $\eta _{1}$ and $\dot{\eta}_{1}$ from
the equations $F=0$. Thus, we get $F=0\rightarrow \bar{F}_{A}\left( \eta
_{2},\dot{\eta}_{2}\right) =0$. Then one has to select independent w.r.t. $%
\dot{\eta}_{2}$ functions $\bar{F}_{A_{/1}}$ . At the same time one finds
new constraints $\bar{F}_{A_{/2}}\left( \eta _{2}\right) =0$ and so on (see
the Subsect. ''First order equations'').

We see that the Dirac procedure differs from our reduction procedure.
Indeed, as was mentioned above, in the Dirac procedure one excludes all the
derivatives $\dot{\eta}$ by the help of the equations $F=0$ from the
consistency conditions $\dot{\Phi}^{\left( 1\right) }=0.$ Thus, one gets
equations for the Lagrange multipliers and new constraints. Besides, one of
the aim of the Dirac procedure is to maintain the canonical Hamiltonian
structure of the equations $F=0.$ The possibility of the Dirac reduction is
due to the specific structure of the equations (\ref{r.1}). Namely, here the
consistency conditions never involve $\dot{\lambda}\,$ and \textrm{%
rank\thinspace }$\partial F/\partial \dot{\eta}=\left[ F\right] =\left[ \eta %
\right] $.

Besides, one ought to mention the work \cite{FadJa88} where it was proposed
an alternative (to the Dirac procedure) way of reducing the equations of
motion for theories with actions of the form $S=\int \left[ \varphi
_{A}\left( \eta \right) \dot{\eta}^{A}-V\left( \eta \right) \right] dt$ .
One can verify that, in fact, the procedure of that work, in a part (the
procedure does not reveal the gauge identities), is similar to our reduction
procedure in the case of the first order equations (see Sect. IV).

However, the reduction procedure proposed in the present article is
formulated for a wider class of Lagrangian systems (differential equations).
It does not need the introduction of new variables such as momenta and
Lagrange multipliers, and it is defined in the framework of the initial
Lagrangian formulation. Moreover, its aim is twofold: to reduce ELE to their
canonical form and to reveal the gauge identities between the ELE equations.

The consideration in the present article is restricted by finite-dimensional
systems. Its application to field theories (theories with infinite number
degrees of freedom) demands additional study. We hope to present the
corresponding formulation in futures publications. However, in simple cases,
one can apply the present reduction procedure with some natural
modifications in the infinite-dimensional case. Consider the Maxwell action $%
S=-\left( 1/4\right) \int \mathcal{F}_{\mu \nu }\mathcal{F}^{\mu \nu }dx\,,$ 
$\;\mathcal{F}_{\mu \nu }=\partial _{\mu }A_{\nu }-\partial _{\nu }A_{\mu }$
as a common example of a singular field theory. The ELE read:%
\begin{eqnarray}
F^{i} &=&\frac{\partial S}{\partial A_{i}}=\partial _{\nu }\mathcal{F}^{i\nu
}=\ddot{A}^{i}+\partial _{i}\dot{A}^{0}-\bigtriangleup A^{i}+\partial
_{i}\varphi =0\,,  \label{r.2} \\
F^{0} &=&-\frac{\partial S}{\partial A^{0}}=\partial _{\nu }\mathcal{F}^{\nu
0}=\dot{\varphi}+\bigtriangleup A^{0}\,=0\,,\;\varphi =\partial _{k}A^{k}\,.
\label{r.3}
\end{eqnarray}%
The equation $F_{0}=0$ is a constraint. Following the reduction procedure,
we have to consider the set $F_{i}=0\,,\;\dot{F}_{0}=0\,.$ The Jacobi matrix 
$\partial F^{\mu }/\partial \ddot{A}_{\nu }$ has the constant rank $3.$ We
can, for example, select the equations (\ref{r.2}) as independent with
respect to the derivatives $\ddot{A}^{i}$. The equation $\dot{F}_{0}=0$%
\thinspace is their consequence. No more constraints appear. Now we exclude $%
A^{0}$ and $\dot{A}^{0}$ from (\ref{r.2}) by the help of (\ref{r.3}). That
creates the equivalence 
\begin{equation}
\left( 
\begin{array}{c}
F^{i} \\ 
F^{0}%
\end{array}%
\right) =\left( 
\begin{array}{cc}
\delta _{k}^{i} & -\frac{\partial _{i}\partial _{0}}{\bigtriangleup } \\ 
0 & 1%
\end{array}%
\right) \left( 
\begin{array}{c}
\bar{F}^{k} \\ 
F^{0}%
\end{array}%
\right) \,,\;\;\;\bar{F}^{k}=\left. F^{k}\right| _{F^{0}=0}=\square \left(
A^{k}+\partial _{k}\varphi \right) \,.  \label{r.4}
\end{equation}%
Now we discover that the functions $\bar{F}^{k}$ are dependent, $\partial
_{k}\bar{F}^{k}\equiv 0.$ In our terms that reads, for example, as the
following equivalence%
\begin{equation}
\left( 
\begin{array}{c}
\bar{F}^{1} \\ 
\bar{F}^{2} \\ 
\bar{F}^{3}%
\end{array}%
\right) =\left( 
\begin{array}{ccc}
1 & 0 & 0 \\ 
0 & 1 & 0 \\ 
-\partial _{3}^{-1}\partial _{1} & -\partial _{3}^{-1}\partial _{2} & 1%
\end{array}%
\right) \left( 
\begin{array}{c}
\bar{F}^{1} \\ 
\bar{F}^{2} \\ 
0%
\end{array}%
\right) \,.  \label{r.5}
\end{equation}%
The equations $\bar{F}^{1}=0,\;\bar{F}^{2}=0,\;F^{0}=0$ present one of the
canonical forms of the Maxwell equations. The identity that follows from the
presence of the zero in the right column of (\ref{r.5}) reads as $\partial
_{\mu }F^{\mu }=0$ in terms of the initial functions $F^{\mu }$ and implies
the invariance of the Maxwell action under gradient gauge transformations.

\begin{acknowledgement}
B. Geyer thanks the foundations FAPESP and DAAD for support and Institute of
Physics USP for hospitality; D. Gitman is grateful to the foundations
FAPESP, CNPq, DAAD for support as well as to the Lebedev Physics Institute
(Moscow) and to the Institute of Theoretical Physics (University of Leipzig)
for hospitality; I. Tyutin thanks INTAS 00-00262, RFBR 02-02-16944,
00-15-96566 for partial support.
\end{acknowledgement}

\section{Appendix}

Here we present three Lemmas which are used in the reduction procedure to
justify equivalence of equations and LF. In this respect it is useful to
recall here the relevant definitions from the Sect. II . Namely:

Two sets of equations, $F_{A}\left( q^{\left[ l\right] }\right) =0$ and $%
f_{\alpha }\left( q^{\left[ l\right] }\right) =0$ are equivalent $%
F=0\Longleftrightarrow f=0$ whenever they have the same set of solutions. If
two sets of LF $F_{A}\left( q^{\left[ l\right] }\right) $ and $\chi
_{A}\left( q^{\left[ l\right] }\right) $ , $\ \left[ F\right] =\left[ \chi %
\right] \,,$ are related by some LO $\hat{U}$ and $\hat{V}$ as $F=\hat{U}%
\chi \,$,\ $\chi =\hat{V}F\,$,\ $\hat{U}\hat{V}=1\,,$ then we call such LF
equivalent and denote this fact as: $F\sim \chi $ . In this case the
corresponding equations are strongly equivalent.

\begin{equation*}
{\Large Lemma\ 1}
\end{equation*}%
Let a set of equations 
\begin{equation}
\Phi _{\mu }\left( x,y^{\left[ l\right] }\right) =0\,,\;F_{a}\left(
x,y\right) =0\,,\;x=\left( x^{\mu }\right) ,\;y=\left( y^{a}\right) \,,
\label{L.3}
\end{equation}%
be given, where $\Phi $ are some LF. And let $\det \left. \partial
F_{a}/\partial y^{b}\right| _{x_{0},y_{0}}\neq 0,\;$where the consideration
point $\left( x_{0},y_{0}\right) $ is on shell. Then:

a) The equations $F_{a}\left( x,y\right) =0$ can be solved w.r.t. $y$ as: $%
y^{a}=\varphi ^{a}(x)\,,$ where $\varphi ^{a}(x)$ are some single-valued
functions of $x$ in the vicinity of the point $x_{0}$\thinspace . In other
words, there is the equivalence 
\begin{equation}
F_{a}\left( x,y\right) \sim y^{a}-\varphi ^{a}(x)\,,  \label{L.2}
\end{equation}%
which implies the strong equivalence between the equations $F_{a}\left(
x,y\right) =0$ and $y^{a}=\varphi ^{a}(x)\,$.

b) The \ following equivalence between the LF holds true:%
\begin{equation}
\left( 
\begin{array}{c}
\Phi _{\mu }\left( x,y^{\left[ l\right] }\right) \\ 
F_{a}\left( x,y\right)%
\end{array}%
\right) \sim \left( 
\begin{array}{c}
\bar{\Phi}_{\mu }\left( x^{\left[ l\right] }\right) \\ 
y^{a}-\varphi ^{a}(x)%
\end{array}%
\right) \,,\;\;\bar{\Phi}_{\mu }=\left. \Phi _{\mu }\right| _{y^{\left[ l%
\right] }=\varphi ^{\left[ l\right] }}\;\,.  \label{l.5}
\end{equation}

The first statement is, in fact, the well-known implicit function theorem %
\cite{Dieud68}. Taking into account (\ref{L.2}), we have: $F_{a}\left(
x,y\right) =u_{ab}\left( y^{a}-\varphi ^{a}(x)\right) \,,\;\left. \det
\,u\right| _{x_{0},y_{0}}\neq 0.$ On the other side one can write $\Phi
_{\mu }=\bar{\Phi}_{\mu }+\hat{V}_{\mu a}\left[ y^{a}-\varphi ^{a}(x)\right]
\,,$ where $\hat{V}_{Aa}$ is a LO. Thus, 
\begin{equation}
\left( 
\begin{array}{c}
\Phi \\ 
F%
\end{array}%
\right) =\hat{U}\left( 
\begin{array}{c}
\bar{\Phi} \\ 
y-\varphi%
\end{array}%
\right) ,\;\;\hat{U}=\left( 
\begin{array}{cc}
1 & \hat{V} \\ 
0 & u%
\end{array}%
\right) ,\;\hat{U}^{-1}=\left( 
\begin{array}{cc}
1 & -\hat{V} \\ 
0 & u^{-1}%
\end{array}%
\right) ,  \label{L.1a}
\end{equation}%
and the equivalence (\ref{l.5}) is justified.%
\begin{equation*}
{\Large Lemma\ 2}
\end{equation*}%
Let a set of equations 
\begin{equation*}
F_{A}\left( q,z\right) =0\,,\;q=\left( q^{a}\right) ,\;z=\left( z^{i}\right)
,\;\;A=1,...,m,\;a=1,...,n\,,\;i=1,...,l\,,
\end{equation*}%
be given. And let the Jacobi matrix $\partial F_{A}/\partial q^{a}$ have a
constant rank in a vicinity $D_{0}$\ of the consideration point $\left(
q_{0}\,,z_{0}\right) $, which is on shell ($F_{A}\left( q_{0}\,,z_{0}\right)
=0$), 
\begin{equation}
\mathrm{rank\,}\left. \frac{\partial F_{A}}{\partial q^{a}}\right| _{q,z\in
D_{0}}=r\,.  \label{l.2}
\end{equation}%
Then there exists an equivalence 
\begin{equation}
F_{A}\sim \bar{F}_{A}=\left( 
\begin{array}{c}
y^{\mu }-\varphi ^{\mu }(x,z) \\ 
\Omega _{G}\left( z\right)%
\end{array}%
\right) \,,\;\;q^{a}=\left( x^{g},y^{\mu }\right) ,\;\;A=\left( \mu
,G\right) ,\;\left[ \mu \right] =r\,.  \label{l.2a}
\end{equation}

We begin the proof with the remark that due to (\ref{l.2}), there exists a
division of the indices $A=\left( \mu ,G\right) $,\ $a=\left( \mu
\,,g\right) \,$,\ $\left[ \mu \right] =r\,$,\ $q^{a}=\left( x^{g},y^{\mu
}\right) \,$,\ such that 
\begin{equation}
\det \left. \frac{\partial F_{\mu }}{\partial y^{\nu }}\right|
_{q_{0},z_{0}}\neq 0\,\,.  \label{l.3}
\end{equation}%
Then by virtue of the Lemma 1 we can write 
\begin{equation}
F_{\mu }=u_{\mu \nu }f^{\nu },\;\;f^{\nu }=y^{\nu }-\varphi ^{\nu
}(x,z),\;\det \left. u\right| _{q_{0},z_{0}}\neq 0\,.  \label{l.4}
\end{equation}%
Let us present the functions $F_{G}$ in the form $F_{G}\left( x,y,z\right)
=\Omega _{G}\left( x,z\right) +\Pi _{G\mu }f^{\mu }\left( x,y,z\right) ,$
where $\;\Omega _{G}\left( x,z\right) =\left. F_{G}\right| _{y=\varphi
\left( z,x\right) }\,,$ such that $\Omega _{G}\left( x_{0},z_{0}\right)
=0\,. $ Then 
\begin{equation}
F_{A}=\left( 
\begin{array}{c}
F_{\mu } \\ 
F_{G}%
\end{array}%
\right) =U_{AB}\chi _{B}\,,\;\;\chi _{B}=\left( 
\begin{array}{c}
f^{\mu } \\ 
\Omega _{G}%
\end{array}%
\right) \,,\;\;U=\left( 
\begin{array}{cc}
u & 0 \\ 
\Pi & 1%
\end{array}%
\right) \,,\;\det \left. U\right| _{q_{0},z_{0}}\neq 0\,.  \label{l.6}
\end{equation}%
In virtue of (\ref{l.2}) and (\ref{l.6}) 
\begin{equation}
\mathrm{rank\,}\left. \frac{\partial \chi _{A}}{\partial q^{a}}\right|
_{q,z\in D_{0}}=r\,.  \label{l.7}
\end{equation}%
The Jacobi matrix $\partial \chi _{A}/\partial q^{a}$\thinspace has the
following structure: 
\begin{equation*}
\frac{\partial \chi _{A}}{\partial q^{a}}=\frac{\partial \left( f^{\mu
},\Omega _{G}\right) }{\partial \left( y^{\nu },x^{g}\right) }=\left( 
\begin{array}{cc}
\delta _{\nu }^{\mu } & -\partial \varphi ^{\mu }/\partial x^{g} \\ 
0 & \partial \Omega _{G}/\partial x^{g}%
\end{array}%
\right) \,.
\end{equation*}%
Therefore,%
\begin{equation}
\mathrm{rank}\,\left. \frac{\partial \Omega _{G}}{\partial x^{g}}\right|
_{x\in D_{0}}=0\Longrightarrow \left. \frac{\partial \Omega _{G}}{\partial
x^{g}}\right| _{x,z\in D_{0}}=0\,.  \label{l.7a}
\end{equation}%
Eq. (\ref{l.7a}), together with the relation $\Omega _{G}\left(
x_{0},z_{0}\right) =0$, implies 
\begin{equation*}
\left. \Omega _{G}\right| _{x,z\in D_{0}}\,=\Omega _{G}\left( z\right)
\,,\;\;\Omega _{G}\left( z_{0}\right) =0\,.
\end{equation*}%
Finally, we may write 
\begin{equation}
F_{A}=U_{AB}\chi _{B}\,,\;\;\chi _{B}=\left( 
\begin{array}{c}
f^{\mu }\left( x,y,z\right) \\ 
\Omega _{G}\left( z\right)%
\end{array}%
\right) \,,\;\;\det \left. U\right| _{q_{0},z_{0}}\neq 0\,,  \label{l.11a}
\end{equation}%
Thus, the equivalence (\ref{l.2a}) is justified.%
\begin{equation*}
{\Large Lemma\;3}
\end{equation*}%
As a consequence of the Lemma 2 the following Lemma holds true:

Let a set of equations%
\begin{equation*}
F_{A}\left( q^{a}\right) =0\,,\;A=1,...,m,\;a=1,...,n\,,
\end{equation*}%
be given. And let the Jacobi matrix $\partial F_{A}/\partial q^{a}$ have a
constant rank in a vicinity $D_{0}$\ of the consideration point $q_{0}$
which is on shell ($F\left( q_{0}\right) =0$), 
\begin{equation*}
\mathrm{rank\,}\left. \frac{\partial F_{A}}{\partial q^{a}}\right| _{q\in
D_{0}}=r\,.
\end{equation*}%
Then there exists an equivalence%
\begin{equation}
F_{A}\sim \bar{F}_{A}=\left( 
\begin{array}{c}
y^{\mu }-\varphi ^{\mu }(x) \\ 
0_{G}%
\end{array}%
\right) \,,\;\;A=\left( \mu ,G\right) ,\;0_{G}\equiv 0\,\;\forall G\,,\;%
\left[ \mu \right] =r\,.  \label{l.12}
\end{equation}

The proof of this Lemma follows the one of the Lemma 2 if one selects there $%
z=z_{0}.$


\begin{thebibliography}{99}
\bibitem{Dirac64} P.A.M. Dirac, \emph{Lectures on Quantum Mechanics,}
(Belfer Graduate School of Science, Yeshiva University, New York 1964)

\bibitem{GitTy90} D.M. Gitman and I.V. Tyutin, \emph{Quantization of Fields
with Constraints}, (Springer-Verlag, Berlin 1990)

\bibitem{HenTe92} M. Henneaux and C. Teitelboim, \emph{Quantization of Gauge
Systems} (Princeton University Press, Princeton 1992)

\bibitem{GitTy01} D.M. Gitman, and I.V. Tyutin, \emph{Constraint
reorganization consistent with Dirac procedure,} hep-th/0112103; Michael
Marinov Memorial Volume: \emph{Multiple Facets of Quantization and
Supersymmetry}, (World Publishing, Singapore 2002)

\bibitem{BorTy98} V.A. Borochov and I.V. Tyutin, Yadernaya Fizika, \textbf{%
61 }(1998) 1715 (Physics of Atomic Nuclei, \textbf{61 }(1998) 1603); ibid 
\textbf{62} (1999) 1137 (Physics of Atomic Nuclei, \textbf{62} (1999) 1070)

\bibitem{BV} I.A. Batalin and G.A. Vilkovisky, Phys. Lett. \textbf{B102}
(1981) 27; Phys. Rev. \textbf{D28} (1983) 2567

\bibitem{GitTy83} D.M. Gitman and I.V. Tyutin, \emph{Canonical quantization
of singular theories}, Izw. VUZov Fizika \textbf{5} (1983) 3 (Sov. Phys.
Journ. \textbf{5} (1983) 423)

\bibitem{GitTy87} D.M. Gitman and I.V. Tyutin, \emph{The structure of gauge
theories in the Lagrangian and Hamiltonian formalisms}, In \emph{Quantum
field theory and quantum statistics} v. I, pp. 143--164, Ed. by Batalin,
Isham and Vilkovisky (Adam Hilger, Bristol 1987)

\bibitem{Kuper79} P. Olver, \emph{Applications of Lie groups to differential
equations}, Graduated Texts in Mathematics, \textbf{107} (Springer-Verlag,
Berlin 1986); B.A. Kupershmidt, \emph{Geometry of jet bundles and structure
of Lagrangian and Hamiltonian formalism, }162-218, Lecture Notes in
Mathematics \textbf{775} (Springer-Verlag, Berlin 1980); I.S. Krasilshchik,
V.V. Lychagin, and A.M. Vinogradov, \emph{Geometry of jet spaces and
nonlinear partial differential equations}, Adv. Studies Contemp. Math.,
(Gordon and Breach, New York 1986)

\bibitem{GitTy02} D.M. Gitman, and I.V. Tyutin,\ Nucl. Phys. \textbf{B630}
(2002) 509

\bibitem{Gantm59} F.R. Gantmacher, \emph{The theory of matrices}, Vol. 1
(Chelsea Publ. Co., New York 1959); \emph{Matrizentheorie}, (Dt. Verlag
Wiss., Berlin 1986)

\bibitem{GitTyL85} D.M. Gitman, S.L. Lyakhovich, and I.V. Tyutin, Izw. VUZov
Fizika \textbf{8} (1983) 61 (Sov. Phys. Journ. \textbf{8} (1983) 730)

\bibitem{Dieud68} J. Dieudonn\'{e}, \emph{Foundations of Modern Analysis, }%
(Academic Press, NY, London 1968); G.M. Fichtenholz, \emph{Differential- und
Integralrechnung}, Vol. I (Dt. Verlag Wiss., Berlin 1989)

\bibitem{FadJa88} L. Faddeev and R. Jackiw, Phys. Rev. Lett. \textbf{60}
(1988) 1692
\end{thebibliography}
\end{document}